\documentclass[twocolumn]{aastex631}

\usepackage{amsmath}
\usepackage{comment}
\usepackage{makecell}

\newcommand{\pfrac}[2]{\left( \frac{#1}{#2} \right)}

\newcommand{\rtidal}{r_{\rm t}}

\newcommand{\MBH}{M_{\bullet}}
\newcommand{\rhod}{\rho_{\rm d}}
\newcommand{\Hd}{H_{\rm d}}
\newcommand{\Sigmad}{\Sigma_{\rm d}}
\newcommand{\kb}{k_{\rm B}}

\newcommand{\dotMEdd}{\dot{M}_{\rm Edd}}

\submitjournal{ApJ}

\begin{document}

\title{QPEs from EMRI Debris Streams Impacting Accretion Disks in Galactic Nuclei}

\correspondingauthor{Itai Linial} \email{il2432@columbia.edu}

\author[0000-0002-8304-1988]{Itai Linial}
\affil{Department of Physics and Columbia Astrophysics Laboratory, Columbia University, New York, NY 10027, USA}

\author[0000-0002-4670-7509]{Brian D.~Metzger}
\affil{Department of Physics and Columbia Astrophysics Laboratory, Columbia University, New York, NY 10027, USA}
\affil{Center for Computational Astrophysics, Flatiron Institute, 162 5th Ave, New York, NY 10010, USA} 

\author[0000-0001-9185-5044]{Eliot Quataert}
\affiliation{Department of Astrophysical Sciences, Princeton University, Peyton Hall, Princeton, NJ 08540, USA}

\begin{abstract}
Quasi-periodic eruption (QPE) sources in galactic nuclei are often associated with a stellar object orbiting a supermassive black hole with hours-days period, brought in as an extreme mass-ratio inspiral (EMRI). In the presence of an accretion disk, repeated star-disk collisions lead to ablation of a small fraction of the stellar mass during each disk passage. We analytically track stellar debris as it is tidally stretched outside the EMRI's Hill sphere, forming an elongated, dilute stream which collides with the disk half an orbit after the last star-disk encounter. For orbital periods $\gtrsim 12$ hr, the dilute stream is deflected at the disk surface by a strong shock, rather than penetrating it. Due to their low optical depth and prolonged interaction time, radiation from the shocked streams typically dominates over emission from shocked disk gas directly impacted by the star. We find that: (1) QPE flare durations reflect the stream-disk collision timescale; (2) Flare luminosities of $10^{42-43}$ erg/s, consistent with observed QPEs, are robustly produced; (3) Soft X-ray flares with temperatures of ${\sim}$100 eV arise when the stream mass is sufficient to sustain a radiation mediated shock at the collision interface. Higher mass streams yield softer flares, typically outshone by the disk, while lower mass streams result in collisionless shocks, which likely produce fainter and harder flares. We discuss observational implications of the temporal evolution of the underlying disk, assuming it is the remnant of a prior tidal disruption event in the same galaxy.
\end{abstract}


\section{Introduction} \label{sec:Intro}

Quasi-Periodic Eruptions (QPEs) are a class of Repeating Nuclear Transients (RNTs), characterized by soft X-ray flares that repeat quasi-periodically on timescales of hours to days, observed in nuclei of nearby ($\lesssim 500 \, \rm Mpc$) galaxies. First discovered in 2019 \citep[GSN 069;][]{Miniutti_2019}, QPE sources have now been found in nearly a dozen galaxies \citep{Giustini_2019,Arcodia_2021,Arcodia_2024_3&4,Arcodia_2025_5,Chakraborty_2021,Chakraborty_25_22upj,Quintin_23,Nicholl_24,Bykov_2025,Hernandez_Garcia_25_Ansky} hosting supermassive black holes (SMBHs) of masses $\MBH \simeq 10^5 - 10^7 \, \rm M_\odot$ at their nuclei \citep[e.g.,][]{Wevers_2022}. 

The range of observed QPE recurrence times, $\left< P_{\rm QPE} \right>$, now spans nearly 2 orders of magnitude, from as short as $\sim 2.4 \, \rm hr$ (eRO-QPE2, \citealt{Arcodia_2021,Arcodia_Linial_24}) and up to multiple days\footnote{The RNT source Swift J0230 \citep{Evans_2023,Guolo_2024} with recurrence time of weeks-month, is not classified as a bona-fide QPE, in part due to its atypical luminosity-temperature evolution.} ($2 \, \rm d$ - AT 2019qiz, \citealt{Nicholl_24}, $\sim 3 \, \rm d$ - AT 2022upj, \citealt{Chakraborty_25_22upj} and $5-10 \, \rm d$ - ``\textit{Ansky}'', \citealt{Hernandez_Garcia_25_Ansky}). QPE flares show a soft, quasi-thermal X-ray spectrum with blackbody temperatures $\kb T_{\rm obs} \approx 100-200 \, {\rm eV}$, hotter than the softer quiescent emission observed between eruptions, $\kb T_{\rm Q} \approx 50-70 \, {\rm eV}$. The bolometric luminosities of QPE flares are approximately $L_{\rm pk} \approx 10^{42-43} \, \rm erg \, s^{-1}$, and their duration $\Delta t_{\rm QPE}$ is typically $\sim$10-20\% of their mean recurrence time \citep[e.g.,][]{Hernandez_Garcia_25_Ansky}.

Several theoretical models for QPEs invoke a star or a compact object, brought onto a tight orbit around the central SMBH as an Extreme Mass Ratio Inspiral (EMRI, \citealt{Linial_Sari_2017})\footnote{The term ``EMRI'' often refers to the mHz gravitational wave (GW) signal produced by the inspiral of a compact object approaching the horizon of an SMBH \citep[e.g.,][]{Amaro_Seoane_2018} -- a key target of future space-based GW detectors. In the context of this paper, EMRI refers more loosely to any stellar object orbiting the SMBH at tens-hundreds of gravitational radii, where the GW luminosity is rather small.}, which provides a natural ``clock'' for setting the QPE periodicity \citep{Zalamea_2010,King_2020,Xian_2021,Sukova_2021,Metzger_2021,Krolik_Linial_2022,Linial_Sari_2023,Lu_Quataert_23,Linial_Metzger_23,Franchini_23,Tagawa_23,Vurm_25}. In some of these models \citep{Linial_Metzger_23,Franchini_23,Tagawa_23,Vurm_25}, QPE-like flares are produced by repeated collisions between an EMRI and a gaseous accretion disk feeding the SMBH. \cite{Linial_Metzger_23} showed that many of the observed properties of short-period QPEs (including their period, luminosity, temperature, pattern of alternating long/short recurrence as well as their occurrence rate in galactic nuclei) can be attributed to a main-sequence star on a mildly eccentric, inclined orbit, that shock-heats the intercepted disk material as it passes through the disk, twice per orbit. 

In most QPE models, the soft X-ray quiescence is attributed to emission from the inner annuli of the underlying accretion disk, accreting at a fraction of the Eddington rate. This disk may originate from: (1) a long-lived AGN accretion flow extending to large radii \citep[e.g.,][]{Xian_2021,Tagawa_23}; (2) Roche lobe overflow from the EMRI itself \citep[e.g.,][]{Linial_Sari_2017,Linial_Sari_2023,Lu_Quataert_23}; or, (3) the tidal disruption of a \textit{second star} approaching the SMBH on a nearly radial orbit (a Tidal Disruption Event, TDE). \cite{Linial_Metzger_23} have demonstrated that the latter scenario (``EMRI+TDE=QPE'') is a natural consequence of the formation rates of EMRIs and TDEs \citep[e.g.,][]{Linial_Sari_2023}, and predicted that $\sim$10\% of TDEs are expected to occur when a tight EMRI is already present around the SMBH, potentially forming a QPE source as collisions between the EMRI and the TDE disk ensue.

Recent studies have advanced key theoretical aspects of the emerging EMRI+disk paradigm. \cite{Vurm_25} performed radiation transport Monte Carlo calculations to model photon production and Comptonization in the ejecta shocked by the star's passage, highlighting the disk conditions and orbiter properties conducive to QPE-like emission. \cite{Yao_2025} used numerical hydrodynamic simulations to study the ablation of main-sequence stars as they repeatedly traverse the disk midplane constraining the lifetime of EMRIs before most of their mass is stripped. They also suggested that collisions between the ablated debris and the disk could dominate the observed emission in QPEs (motivating the scenario we investigate here). \cite{Linial_Metzger_24b} explored the coupled evolution of the EMRI and the disk over timescales of years-decades, accounting for the energy and mass deposited in the disk from the ablated stellar material. Axisymmetric GR-hydrodynamic simulations \citep{Tsz-Lok_Lam_25_BH_simulation} of a stellar-mass black hole traversing an accretion disk have been used to study the resulting outflow and accretion of disk material onto the smaller black hole. Radiation-hydrodynamics simulations of the high Mach number collision between a main-sequence star and a disk \citep{Huang_2025}, provide for the first time synthetic lightcurves and time-dependent spectra that account for much of the relevant physics, facilitating  detailed comparison of theory and QPE observations.

Ongoing observational progress continues to unravel the rich QPE phenomenology and provides important tests for theoretical models. New QPE sources with $\left< P_{\rm QPE} \right> \gtrsim 1\,\rm d$ have been discovered in the aftermath of previously detected TDEs (AT 2019qiz, \citealt{Nicholl_24} and AT 2022upj, \citealt{Chakraborty_25_22upj}), providing strong support to the ``EMRI+TDE=QPE'' hypothesis \citep{Linial_Metzger_23}. Recent followup UV/X-ray observations with \textit{HST} and \textit{XMM-Newton} have characterized the SMBH accretion flow in two QPE sources (GSN 069 and eRO-QPE2), finding in both a compact, TDE-like disk, accreting at a fraction of the Eddington rate, with an outer radius that exceeds the required EMRI semi-major axis \citep{Guolo_2025_GSN069_joint,Guolo_2025_GSN069_time_dependent,Wevers_2025_ero2}. The recently discovered QPE source ``Ansky'' \citep{Hernandez_Garcia_25_Ansky} has appeared following an optical flare interpreted as a ``turn-on AGN'', with X-ray flares repeating every $\left< P_{\rm QPE}\right> \approx 5 \, \rm d$. This source also exhibits varying ionization/absorption features \citep{Chakraborty_25_ionization}, possibly indicative of shock heated material undergoing rapid quasi-spherical expansion, akin to the star-disk collision picture for QPEs \citep{Linial_Metzger_23,Vurm_25}.

With the increasing sample of QPE sources and maturation of theoretical models, the time is ripe for confronting some outstanding observational puzzles and refining theoretical ideas.
\begin{itemize}
    \item \textit{Flare durations scale with recurrence times}. {QPE flare durations, $\Delta t_{\rm QPE}$, increase roughly linearly (or slightly faster) with $\left< P_{\rm QPE} \right>$ \citep{Hernandez_Garcia_25_Ansky}, with a duty cycle $\mathcal{D} \equiv \Delta t_{\rm QPE}/\left< P_{\rm QPE} \right> \approx 0.1-0.2$.} Multi-hour flares in long-period QPEs prove challenging to explain with photon diffusion times of disk material intercepted by a star \citep[e.g.,][]{Yao_2025,Vurm_25,Mummery_25,Guo_25}.
    \item \textit{QPEs appear only in soft X-rays}. {QPEs are detected in soft X-rays ($0.2-1.0 \, \rm keV$), with no coincident variability detected in other bands}. While high-velocity ($\gtrsim0.1 \, c$) collisions with low-density disks may explain X-ray emission due to photon-starved shocks (akin to the early emission in supernova shock-breakout signals, e.g., \citealt{Weaver_76,Nakar_Sari_10,Linial_Metzger_23,Vurm_25}), it remains unclear why do all QPEs exhibit similar $\kb T_{\rm obs} \approx 100-200 \, \rm eV$ and nearly identical spectral evolution patterns (see \citealt{Vurm_25} for further discussion).
    \item {\textit{Total flare energies increase with period}. The radiated energy per flare, $E_{\rm flare} \approx L_{\rm bol} \Delta t_{\rm QPE}$ generally grows with $\left< P_{\rm orb} \right>$, reaching up to $E_{\rm flare} \approx 10^{48} \, \rm erg$ in long-period sources \citep{Hernandez_Garcia_25_Ansky,Chakraborty_25_ionization}. If the emission is arising from shocked disk material, the implied disk masses appear to be unrealistically high \cite[e.g.,][]{Guo_25, Mummery_25}}.
    \item \textit{Onset of QPEs following a TDE}. {At least some QPE sources emerge in the late-time aftermath of a TDE, ``turning-on'' after an optical TDE flare is observed. In the EMRI+disk picture, the TDE disk needs to spread sufficiently to intercept the EMRI orbit, and the QPE flares must be sufficiently bright and hot relative to the quiescent emission \citep{Linial_Metzger_23}. However, the triggering mechanism of QPEs following a TDE is not entirely clear (e.g., the lack of QPE flares in GSN 069 in 2014 and their presence in 2018, and see also \citealt{Miniutti_23,Guolo_2025_GSN069_time_dependent,Nicholl_24}).}
\end{itemize}

In this paper, we study the dynamics of \textit{ablated stellar debris} -- material stripped from a stellar-EMRI undergoing repeated collisions with the underlying disk -- a scenario first discussed in \cite{Yao_2025}. Tidal forces stretch the debris to an elongated stream which returns to impact the rotating disk, half an orbit after the previous EMRI+disk collision (see Fig.~\ref{fig:SchematicOverview} for an illustration). We demonstrate that this interaction inevitably produces an energetic emission component that was underappreciated in earlier EMRI+disk models \citep[e.g.,][]{Xian_2021,Linial_Metzger_23,Franchini_23,Tagawa_23}, but which may dominate the observed signal (as was speculated in \citealt{Yao_2025}). This revised paradigm appears to naturally resolve, at least qualitatively, some of the observational puzzles listed above.

The paper is structured as follows: We introduce the basic model components in \S \ref{sec:Components}, before discussing the dynamics of the ablated stellar debris, its tidal evolution to an elongated stream and its impact with the disk in \S \ref{sec:Debris_and_stream}. Section \S \ref{sec:Observables} is dedicated to the observable signatures of EMRI+disk and debris-stream+disk interactions. We address the radiative efficiency of the process and its implications concerning the emitted temperature in \S \ref{sec:Efficiency}, and address consequences of the assumed disk structure in \S\ref{sec:DiskModels}. We discuss the key results and conclude in Section \S \ref{sec:Discussion}.

\section{Model Components} \label{sec:Components}

\subsection{The stellar-EMRI}
We consider a stellar-EMRI of mass $m_\star$ and radius $R_\star$, orbiting an SMBH of mass $\MBH$ on a circular orbit of period $P_{\rm orb}$, semi-major axis
\begin{multline}
    a_0 = (G\MBH P_{\rm orb}^2 / 4\pi^2)^{1/3} \approx 1 \, {\rm au} \; M_{\bullet,6}^{1/3} P_{\rm orb,8}^{2/3} \\ \approx 95 \, R_{\rm g} \;  (P_{\rm orb,8}/M_{\bullet,6})^{2/3} \,,
\end{multline}
and Keplerian velocity
\begin{equation}
    v_{\rm k} \approx c \, (R_{\rm g}/a_0)^{1/2} \approx 0.1 \, c \; (M_{\bullet,6}/P_{\rm orb,8})^{1/3} \,,
\end{equation}
where $M_{\bullet,6} = \MBH/10^6 \, \rm M_\odot$, $P_{\rm orb,8} = P_{\rm orb}/8 \, \rm hr$ and $R_{\rm g}=G\MBH/c^2$ is the gravitational radius.

We assume that the star is not overflowing its Roche-lobe, and thus $a_0 \gtrsim 2 \, \rtidal$, where $\rtidal = R_\star (\MBH/m_\star)^{1/3}$ is the tidal radius \citep[e.g.,][]{Eggleton_83}. This criterion corresponds to a lower limit on the orbital period
\begin{multline} \label{eq:P_orb_min}
    P_{\rm orb} \gtrsim P_{\rm orb,min}= P_{\rm orb}(2\rtidal) \approx 2^{5/2} \pi \sqrt{R_\star^3 / G m_\star} \\
    \approx 7.9 \, {\rm hr} \; m_1^{0.7} \,,
\end{multline}
where $P_{\rm orb,min}$ is of the order of the star's dynamical timescale, notably independent of $\MBH$. Here we defined $m_1 = m_\star/\rm M_\odot$ and assumed that the star follows the main-sequence mass-radius relation, $R_\star \propto m_\star^{0.8}$. 

As we later discuss, the star-disk interaction produces either one or two QPE-like flares per orbit, i.e., $\left< P_{\rm QPE} \right> \approx P_{\rm orb}$ or $\left< P_{\rm QPE} \right> = P_{\rm orb}/2$, where $\left< P_{\rm QPE} \right>$ is the mean QPE recurrence time. We note that the relatively short period seen in eRO-QPE2 imposes tight constraints on the EMRI mass, if it is indeed a main-sequence star, with $m_\star \lesssim 0.35 \, \rm M_\odot$ (assuming two flares per orbit). Furthermore, if the EMRI orbit is even somewhat eccentric, with $e\gtrsim 0.01$, tidal heating may significantly inflate the star's envelope, rendering it even more susceptible to Roche lobe overflow, implying that mass transfer ensues at a somewhat wider separation, $r_{\rm MT} \approx 4 \, \rtidal$ \citep[e.g.,][]{Linial_Quataert_24b,Yao_2025b}. In the case of eRO-QPE2, satisfying $a_0\gtrsim r_{\rm MT}$ requires a stellar-mass $m_\star \lesssim 0.1 \, \rm M_\odot$ for a main-sequence EMRI.

\subsection{The accretion disk}
We consider an optically thick, geometrically thin accretion disk, characterized by a local scale-height $\Hd(r)$, midplane density $\rhod(r)$ and surface density $\Sigmad(r) \approx \rhod \Hd$. Motivated by the quiescent soft X-ray emission observed in between QPE flares \citep{Miniutti_2019,Giustini_2019,Arcodia_2024_3&4,Linial_Metzger_23}, we consider accretion rates $\dot{M}_{\rm d} = \lambda_{\rm Edd} \dot{M}_{\rm Edd}$, with $10^{-2} \lesssim\lambda_{\rm Edd} \lesssim 1$ and $\dotMEdd = 4\pi G\MBH / (\varepsilon_{\rm acc} \kappa_{\rm es} c) = L_{\rm Edd}/(\varepsilon_{\rm acc} c^2)$ -- the Eddington accretion rate. Here $\varepsilon_{\rm acc} \approx0.1$ is the accretion radiative efficiency and $\kappa_{\rm es} \approx 0.34 \, \rm cm^2 \, g^{-1}$ is the electron scattering opacity. 

At the orbital separations of interest, $a_0\approx\mathcal{O}(10^2) \times R_{\rm g}$, the disk aspect ratio is taken to be $\tilde{h}\equiv\Hd/a_0 \approx 10^{-3}-10^{-1}$, encompassing a range of assumptions regarding the disk's vertical pressure support (e.g., from radiation, gas or magnetic pressure) as well as the dominant sources of opacity. When possible, we remain agnostic regarding the detailed disk structure, angular momentum transport, turbulent stresses and the disk's cooling mechanisms, in order to accommodate theoretical uncertainties associated with these aspects of black hole accretion. We discuss the consequences of more specific assumptions regarding the underlying disk in \S \ref{sec:DiskModels}.

\subsection{Ablation from star-disk encounters}
Assuming the star's orbit is highly inclined with respect to the disk (i.e., $\iota \sim \mathcal{O}(1)$), repeated disk passages deposit energy in the form of shocks within the star's envelope, leading to gradual ablation of stellar material with every passage \citep{Lu_Quataert_23,Linial_Metzger_23,Yao_2025}. High-resolution hydrodynamical simulations by \cite{Yao_2025} find a calibrated prescription for the ablated mass per disk passage\footnote{\cite{Yao_2025} show that the ablation efficiency is substantially smaller during the first few disk passages of an unperturbed main-sequence star. Here we consider the case where the star has undergone several disk passages, with its outer layers somewhat inflated, thereby increasing the ablation efficiency.}
\begin{multline} \label{eq:M_ej_Yao}
    m_{\rm ej,\star} \approx \eta \frac{p_{\rm ram}}{p_\star} m_\star \approx  2\pi \eta \, \rhod \pfrac{m_\star}{\MBH}^{1/3} \frac{\rtidal^4}{a_0} \approx \\
    1.6\times 10^{-5} \, {\rm M_\odot} \, \eta_{0.03} \, \rho_{\rm d,-7}(M_{\bullet,6}/P_{\rm orb,8})^{2/3} R_1^4/m_1
\end{multline}
where $p_{\rm ram} \approx \rhod v_{\rm k}^2 / 2$ is the ram pressure exerted upon the star, $p_\star \approx G m_\star^2 / (4\pi R_\star^4)$ is the mean stellar pressure, and $\eta \approx 0.03$ is the stripping efficiency, obtained for an $n=3/2$ polytrope \citep{Yao_2025}. We define $R_1=R_\star/\rm R_\odot$ and $\eta_{0.03} = \eta/0.03$, and note that for a main-sequence star, $m_{\rm ej,\star}/m_\star \propto m_\star^{1.2}$. The above expression is valid in the regime $H_{\rm d} \gtrsim R_\star$, and could be qualitatively understood by considering the depth within the stellar envelope where the external ram pressure is balanced by the envelope's local hydrostatic pressure. Accounting for two collisions per orbit, the star's mass loss rate due to ablation is roughly
\begin{multline} \label{eq:mdot_abl}
    |\dot{m}_{\rm abl}|\approx \frac{2m_{\rm ej,\star}}{P_{\rm orb}} \approx \\
    0.03 \, {\rm M_\odot \, yr^{-1}} \;
    \eta_{0.03} \, \rho_{\rm d,-7} \,M_{\bullet,6}^{2/3} P_{\rm orb,8}^{-5/3} m_1^{2.2} \approx \\
    1 \;\dotMEdd \;
    \eta_{0.03} \, \rho_{\rm d,-7} \,M_{\bullet,6}^{-1/3} P_{\rm orb,8}^{-5/3} m_1^{2.2} \,,
\end{multline}
and the ablation timescale
\begin{equation}
    \tau_{\rm abl} \approx \frac{m_\star}{|\dot{m}_{\rm abl}|} \approx 30 \, {\rm yr} \; \eta_{0.03}^{-1}\rho_{\rm d,-7}^{-1} m_1^{-1.2} M_{\bullet,6}^{-2/3} P_{\rm orb,8}^{5/3} \,,
\end{equation}
setting an upper limit on the duation of QPE activity, if driven by stellar-EMRIs\footnote{QPE sources have only been observed over a baseline of a few years at this point, and thus their lifetimes are not well constrained from the available data. The current record holder is RXJ J1301 \citep{Giustini_2019,Giustini_2024}, which has been displaying QPE flares for more than 20 years at this point -- setting important constraints on the nature of the EMRI, the ablation process and the QPE lifetime.}. The rate at which mass is stripped from the star could be comparable to, or even exceed the disk's accretion rate, $\dot{M}_{\rm d}$. The coupled evolution of the impacting star and the disk fed by the ablated stellar mass, has been recently studied in \cite{Linial_Metzger_24b} - who found the existence of equilibria states where the disk is steadily fed by stellar ablation (namely, $\dot{M}_{\rm d} \simeq |\dot{m}_{\rm abl}|$) as well as limit-cycles where the disk transitions between states of low and high accretion rates, over timescales of years to decades.

\subsection{Directly impacted disk material}

The amount of disk material directly swept up by the star is governed by its physical cross section, $m_{\rm ej,d} \approx 2\pi R_\star^2 \Sigmad$, where the factor of 2 accounts for the disk's azimuthal flow, perpendicular to the star's trajectory, increasing the amount of mass swept up by the star \citep{Linial_Metzger_23}. It is usually smaller than the stripped stellar mass, $m_{\rm ej,\star}$, with
\begin{multline} \label{eq:m_ej_vs_ejected_disk_mass}
    \frac{m_{\rm ej,\star}}{m_{\rm ej,d}} \approx \eta \pfrac{\MBH}{m_\star}^{2/3} \pfrac{\rtidal}{a_0} \pfrac{R_\star}{H_{\rm d}} \\
    \approx 70 \; \eta_{0.03} P_{\rm orb,8}^{-4/3} \MBH^{1/3} m_1^{0.6} \tilde{h}_{-2}^{-1} \,.
\end{multline}
While the mass $m_{\rm ej,d}$ is shock-accelerated to roughly the local orbital velocity, $v_{\rm k}$ \citep[e.g.,][]{Linial_Metzger_23}, the bulk of the ablated stellar mass, $m_{\rm ej,\star}$, is only marginally gravitationally unbound from the star, ejected with velocities comparable to the star's escape speed $v_{\rm ej,\star} \gtrsim v_{\rm esc,\star}$ \citep{Yao_2025}. As seen in the star's frame, the kinetic/internal energy content of the directly impacted disk material exceeds that of the stellar debris
\begin{multline}
    \left. \frac{E^0_{\rm ej,d}}{E^0_{\rm ej,\star}} \right|_{\rm \star} \approx \frac{m_{\rm ej,d}}{m_{\rm ej,\star}} \pfrac{v_{\rm k}}{v_{\rm esc,\star}}^2 \approx \eta^{-1} \frac{H_{\rm d}}{R_\star} \\
    \approx 70\; \eta_{0.03}^{-1} \tilde{h}_{-2} P_{\rm orb,8}^{2/3} M_{\bullet,6}^{1/3} m_1^{-0.8} \,,
\end{multline}

However, in the frame of the rotating disk, both components ($m_{\rm ej,d}$ and $m_{\rm ej,\star}$) are moving at velocities comparable to $v_{\rm k}$, suggesting that collisions between the \textit{ablated stellar material and the disk} could be energetically dominant, as previously highlighted in \cite{Yao_2025} (e.g., their Figure 9). In the sections that follow we demonstrate that debris-disk collisions can convert a large fraction of the available kinetic energy $m_{\rm ej,\star} v_{\rm k}^2/2$ into radiation, providing a natural mechanism for powering RNTs and QPEs in particular.

\begin{figure*}
    \centering
    \includegraphics[width=\textwidth]{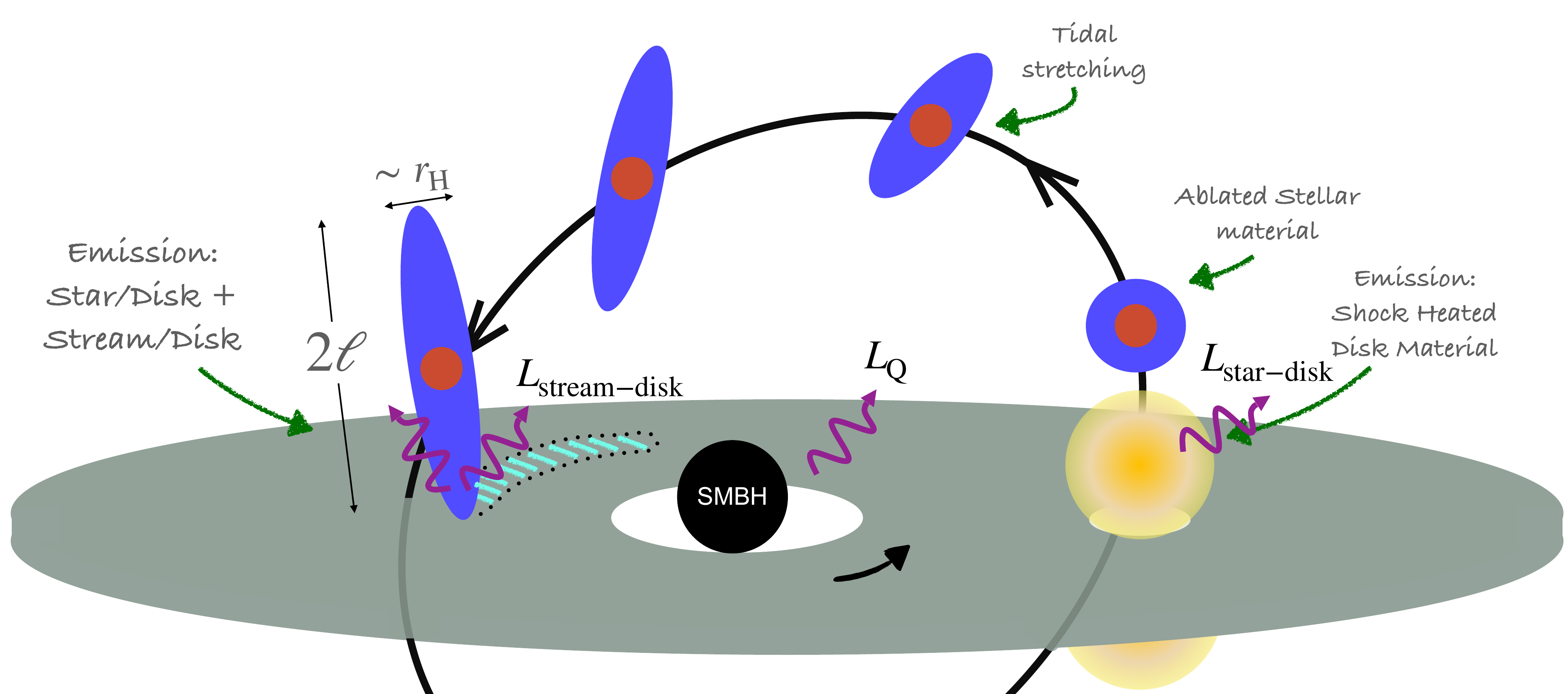}
    \caption{A schematic illustration of the model, involving an SMBH engulfed by a accretion disk, and a star on an inclined, low-eccentricity orbit. Collisions between the star and the disk eject shock heated material above and below the plane of the disk, resulting in emission discussed in \cite{Linial_Metzger_23}. Encounters with the disk also result in ablation, stripping a small fraction of the star's mass with every passage (as studied in \citealt{Yao_2025}). The stellar debris quickly fills the star's Hill sphere, beyond which tidal forces shape it into an elongated stream, impacting a large area of the disk (highlighted on the disk surface) after ${\sim}1/2$ an orbit. The stream is shocked and decelerated following the collision, resulting in a bright flare, with total energy exceeding that arising from the direct star-disk collision.}
    \label{fig:SchematicOverview}
\end{figure*}

\section{Stellar Debris and Stream Formation} \label{sec:Debris_and_stream}

We begin by discussing the evolution of the stellar debris ablated from the star as it repeatedly passes through the disk. Our goal here is to characterize the geometry of previously liberated debris, as it approaches the disk for an impact.

\subsection{Early Times - Within the Hill sphere}

Following disk crossing, ablated material engulfs the star with typical velocity (in the star's frame) $v_{\rm ej,\star} \gtrsim v_{\rm esc,\star}$. We make the simplifying assumption that the ablated material expands homologously outward from the stellar surface, with the majority of the unbound mass, $\sim m_{\rm ej,\star}$, propagating radially at velocity $v_{\rm ej,\star}$. The ejecta expands to fill the star's Hill sphere within {a time $t_{\rm H}$}, which is a small fraction of the orbital period
\begin{multline}
    \frac{t_{\rm H}}{P_{\rm orb}} \approx\frac{r_{\rm H}-R_\star}{v_{\rm ej,\star} P_{\rm orb}} \lesssim \frac{ 1 }{2\pi} \pfrac{\rtidal}{a_{\rm 0}}^{1/2} \left[ 1- \pfrac{\rtidal}{a_0}\right] \\
    \lesssim 0.1 \, (a_0/2\rtidal)^{-1/2} \approx 0.1 \; P_{\rm orb,8}^{-1/3} m_1^{-0.23}\,,
\end{multline}
where $r_{\rm H} \equiv a_0 (m_\star/\MBH)^{1/3}$ is comparable to the star's Hill radius (up to a factor $3^{1/3}$ commonly included in this definition). Slow moving material, marginally bound to the star, remains within the Hill sphere until the following disk encounter at time $t \approx P_{\rm orb}/2$.

The mean ejecta density at time $t_{\rm H}$, relative to the midplane disk density, is approximately
\begin{multline} \label{eq:rho_ej_initial}
    \frac{\bar{\rho}_{{\rm ej},i}}{\rhod} \approx  \frac{m_{\rm ej,\star}}{4\pi \rhod r_{\rm H}^3/3} \approx \frac{3}{2} \eta \pfrac{\MBH}{m_\star}^{2/3} \pfrac{\rtidal}{a_0}^4 \approx \\
    26 \; \eta_{0.03} M_{\bullet,6}^{2/3} m_1^{1.2} P_{\rm orb,8}^{-8/3} \,,
\end{multline}
and its initial optical depth upon filling the Hill sphere
\begin{equation} \label{eq:optical_depth_inital}
    \tau_{{\rm ej},i} \approx \frac{\kappa m_{\rm ej,\star}}{4\pi r_{\rm H}^2} \approx 4\times 10^4 \; \eta_{0.03} \, \rho_{\rm d,-7} \,M_{\bullet,6}^{2/3} m_1^{1.5} P_{\rm orb,8}^{-2} \,.
\end{equation}

It will prove useful to consider the amount of disk mass intercepted by the star's Hill sphere as it passes through the disk
\begin{multline} \label{eq:m_disk_Hill}
    m_{\rm d,H} = m_{\rm ej,d} (r_{\rm H}/R_\star)^2\approx \pi \rhod a_0^3 \pfrac{m_\star}{\MBH}^{2/3} \tilde{h} = \\
    8.8\times 10^{-7} \, {\rm M_\odot} \, \rho_{\rm d,-7}\tilde{h}_{-2} M_{\bullet,6}^{1/3} m_\star^{2/3} P_{\rm orb,8}^{2} \,,
\end{multline}
and in comparison with the ejecta mass, $m_{\rm ej,\star}/m_{\rm d,H} \approx 18 \, \eta_{0.03} P_{\rm orb,8}^{-8/3} M_{\bullet,6}^{1/3} \tilde{h}_{-2}^{-1} R_1^4 m_1^{-5/3}$.

\subsection{Outside the Hill sphere and the following disk encounter}
After evacuating the Hill sphere, the unbound ejecta evolves due to the SMBH's tidal gravity, with subdominant contribution from the star's gravity or pressure gradients\footnote{The stripped material is in approximate virial equilibrium (i.e., comarable amounts of kinetic and internal energy) shortly after disk passage, on scales $R_\star\lesssim r \lesssim 2R_\star$. By the time the ejecta has expanded to fill the Hill sphere ($r_{\rm H} > R_\star$), its internal energy has decreased considerably through adiabatic expansion, and thus we assume a cold gas in the subsequent evolution.}. In this limit, the ejecta evolves at times $t\gtrsim t_{\rm H}$ satisfying Hill's equations -- the linearized equations of motion in a local, co-orbiting frame centered around the star, and neglecting its own gravity
\begin{eqnarray} \label{eq:Hill_x}
    \ddot{x} & = & 3 \Omega_0^2 x + 2\Omega_0 \dot{z} \,,\\
    \ddot{y} & = & -2 \Omega_0 \dot{x} \,, \\ \label{eq:Hill_z}
    \ddot{z} & = & -\Omega_0^2 z \,.
\end{eqnarray}
where $\Omega_0 = \sqrt{G\MBH/a_0^3}$, and the Cartesian coordinates are defined by $\hat{x}$ -- pointed away from the SMBH, $\hat{y}$ -- directed along the star's orbit, and $\hat{z}$ -- parallel to the angular velocity, $\mathbf{\Omega}_0$.

We assume an idealized initial configuration of the ejecta at time $t=t_{\rm H}$, composed of an homologous density profile described by $r_{\rm i}(m)$ and $v_{\rm i}(m) \approx v_{\rm ej,\star} (r_{\rm i}(m)/r_{\rm H})$. In this notation, $r_{\rm i}(m)$ is the initial shell radius enclosing mass $m \lesssim m_{\rm ej,\star}$, such that $r_{\rm i}(m_{\rm ej,\star})\approx r_{\rm H}$. The assumption of homologous expansion implies that all mass shells share the following parameter
\begin{equation}
    C \equiv \frac{v_{\rm i}}{\Omega_0r_{\rm i}} = \frac{v_{\rm ej,\star}}{\Omega_0r_{\rm H}} \approx \sqrt\frac{a_0}{\rtidal} = 1.4 \; P_{\rm orb,8}^{1/3} m_\star^{0.23} \,,
\end{equation}
representing the initial velocity relative to the tidal shear velocity - scaling weakly with orbital period and stellar properties.

Every mass element initially located on a spherical shell at position $r_{\rm i}( \sin{\theta} \cos{\phi},\sin{\theta} \sin{\phi},\cos{\theta})$, with $\theta$ and $\phi$ the polar and azimuthal angles (defined in the same coordinate system centered at the star), takes the following trajectory, given by a straightforward solution of Hill's Eqs.~\ref{eq:Hill_x}-{\ref{eq:Hill_z}
\begin{widetext}
\begin{eqnarray}
    \label{eq:Hill_sol_x}
    x(\tilde{t})/r_{\rm i} &=& [ (4-3 \cos{\tilde{t}}) + C \sin{\tilde{t}} ] \sin{\theta} \cos{\phi}  + 2(1-\cos{\tilde{t}}) C \sin{\theta} \sin{\phi} \,, \\
    y(\tilde{t})/r_{\rm i} &=& [ 6(\sin{\tilde{t}}-\tilde{t}) + 2 (\cos{\tilde{t}}-1) C] \sin{\theta} \cos{\phi} + [ 1 + (4\sin{\tilde{t}}-3\tilde{t}) C ] \sin{\theta} \sin{\phi} \,, \\ \label{eq:Hill_sol_z}
    z({\tilde{t}})/r_{\rm i} &=& \cos{\tilde{t}} \cos{\theta} + C \sin{\tilde{t}} \cos{\theta} \,,
\end{eqnarray}
\end{widetext}
where $\tilde{t} = \Omega_0 (t - t_{\rm H}) \approx \Omega_0 t - 1/C$. These ``free'' solutions to Hill's equations are essentially the ballistic, Keplerian trajectories of ejecta particles, as seen in the frame following the star's circular orbit, taken to first order in $\sqrt{x^2+y^2+z^2}/a_0 \ll 1$. We also note that the solution is sensible only for initial radii comparable to $r_{\rm H}$, which quickly evacuate the Hill sphere, and it is definitely innaccurate for $r_{\rm i} \ll r_{\rm H}$, where initial pressure gradients and the star's gravity are non-negligible.

Remarkably, the surface defined through $x(\tilde{t}),y(\tilde{t}),z(\tilde{t})$, (and parameterized by $\theta,\phi$) evolves from a sphere of radius $r_{\rm i}$ at $\tilde{t}=0$ into a \textit{triaxial ellipsoid} centered at the star, with axes $r_{\rm i} \tilde{R}_1(\tilde{t}), r_{\rm i} \tilde{R}_2(\tilde{t}) , r_{\rm i} \tilde{R}_3(\tilde{t})$. 
The axis $\tilde{R}_3$ coincides with the $\hat{z}$ direction, whereas the axes $\tilde{R}_{1}$ and $\tilde{R}_{2}$ form an angle $\alpha(\tilde{t})$ with the $\hat{x}$ and $\hat{y}$ axes, respectively (Fig.~\ref{fig:FreeSolutionEllipsoid}, top-left panel). The assumption of initial homologous expansion (i.e., uniform $C$ for all shells) is maintained at later times, and thus, all shells evolve in a self-similar manner, retaining their initial ordering.

\begin{figure*}
    \centering
    \includegraphics[width=0.49\textwidth]{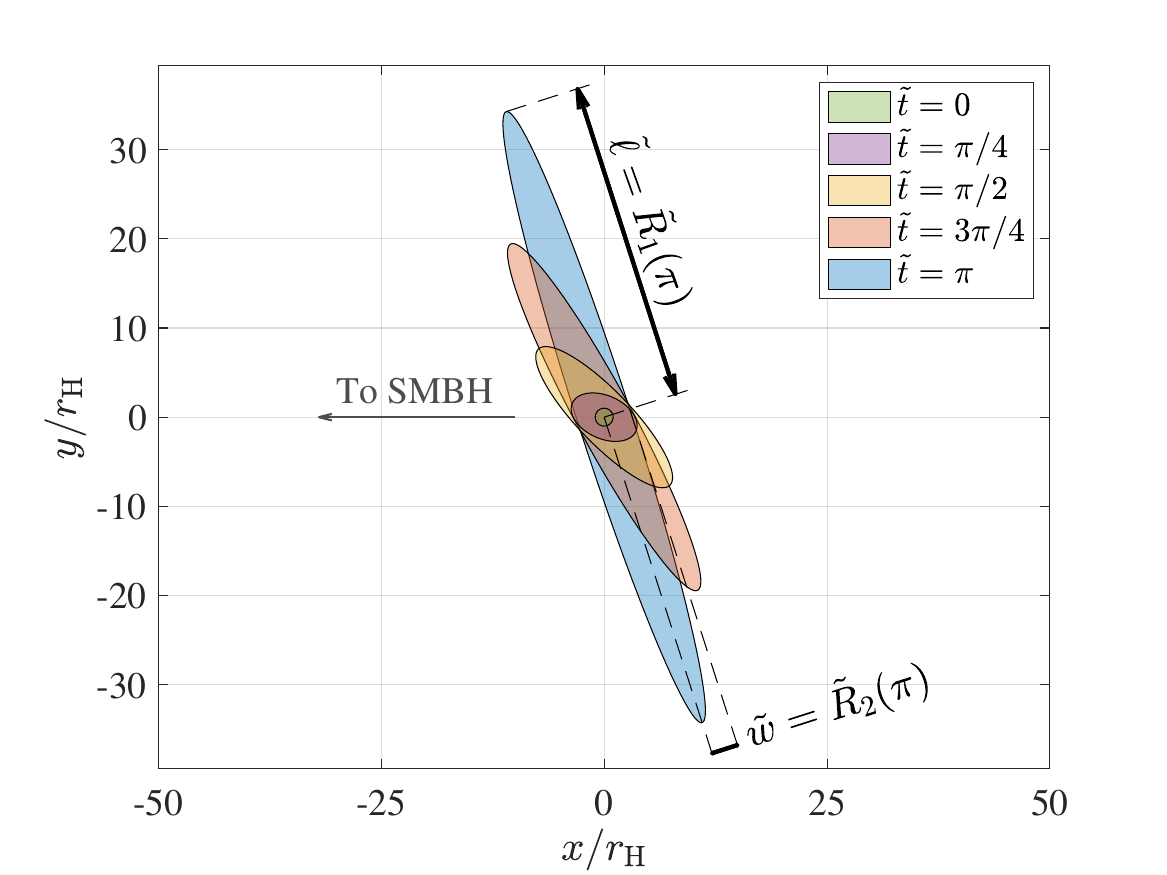}
    \includegraphics[width=0.49\textwidth]{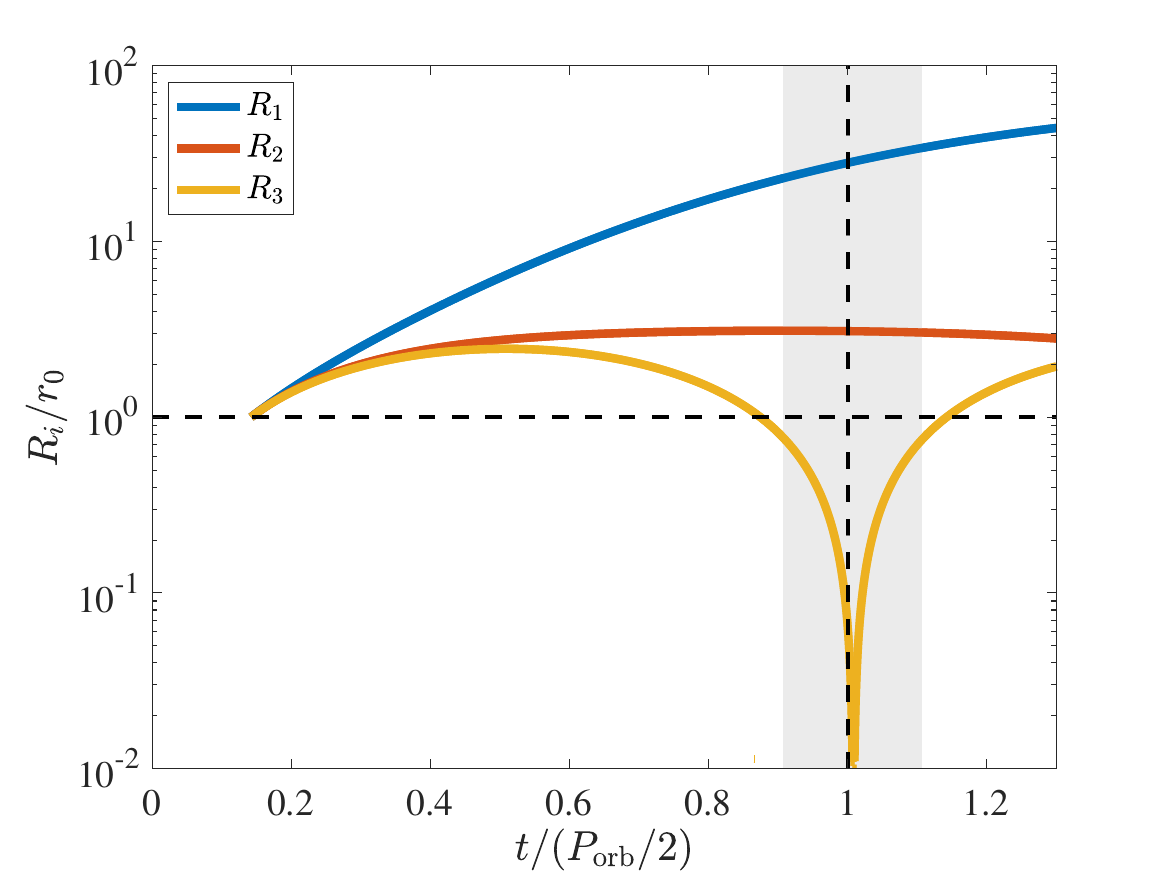}
    \caption{Time evolution of the debris geometry, assuming $a_0/\rtidal = 5$. Left: The ellipsoid projection onto the orbital plane at different times, starting from the Hill sphere at $\tilde{t}=0$, and forming an elongated stream by the time of the next disk encouner, $\tilde{t}\approx\pi$. Right: The ellipsoid primary axis as a function time. At time $t_{\rm c} \approx P_{\rm orb}/2$ (Eq.~\ref{eq:tau_max_compression}), $R_3$, the ellipsoid axis parallel to the $z$ direction approaches 0. Increasing pressure gradients halt the vertical compression, resulting in a ``bounce''. The shaded region shows the star-disk collision interval $(t_1,t_2)$.}
    \label{fig:FreeSolutionEllipsoid}
\end{figure*}

\begin{figure}
    \centering
    \includegraphics[width=\linewidth]{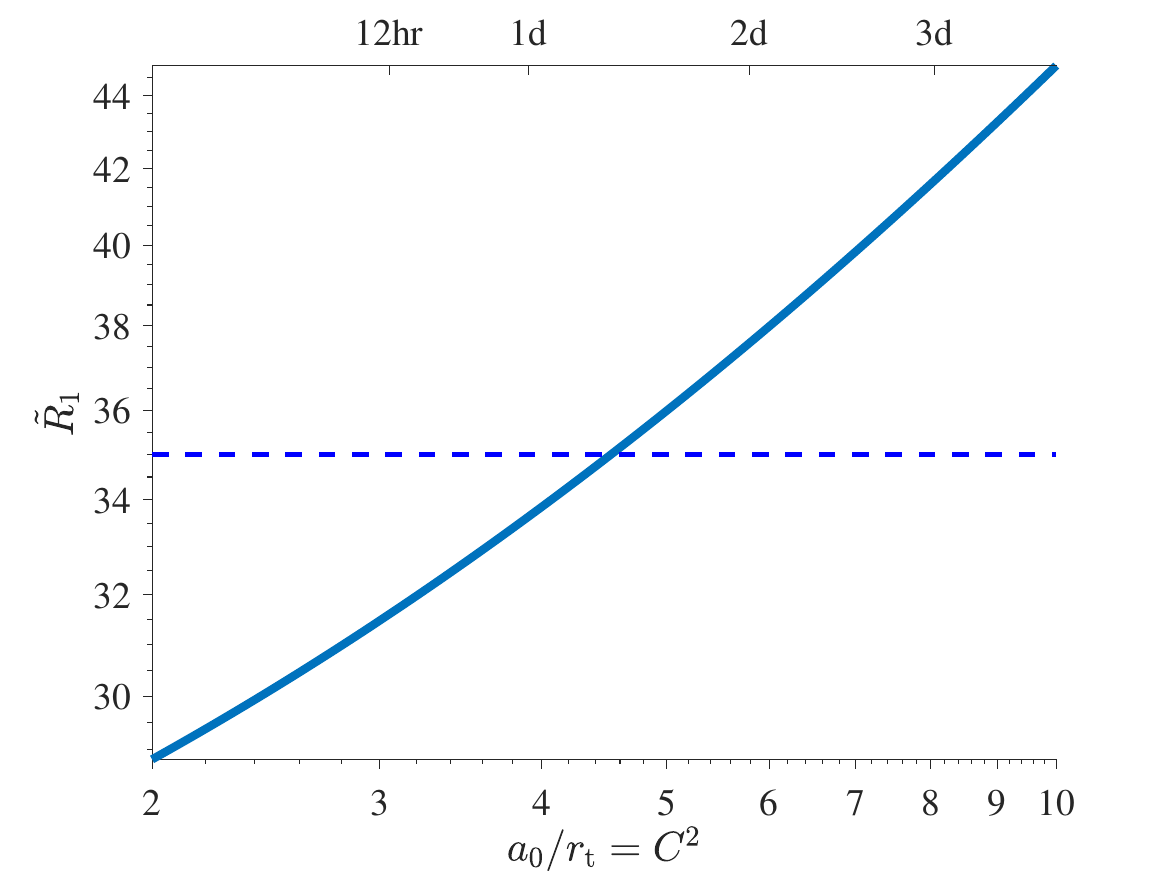}
    \includegraphics[width=\linewidth]{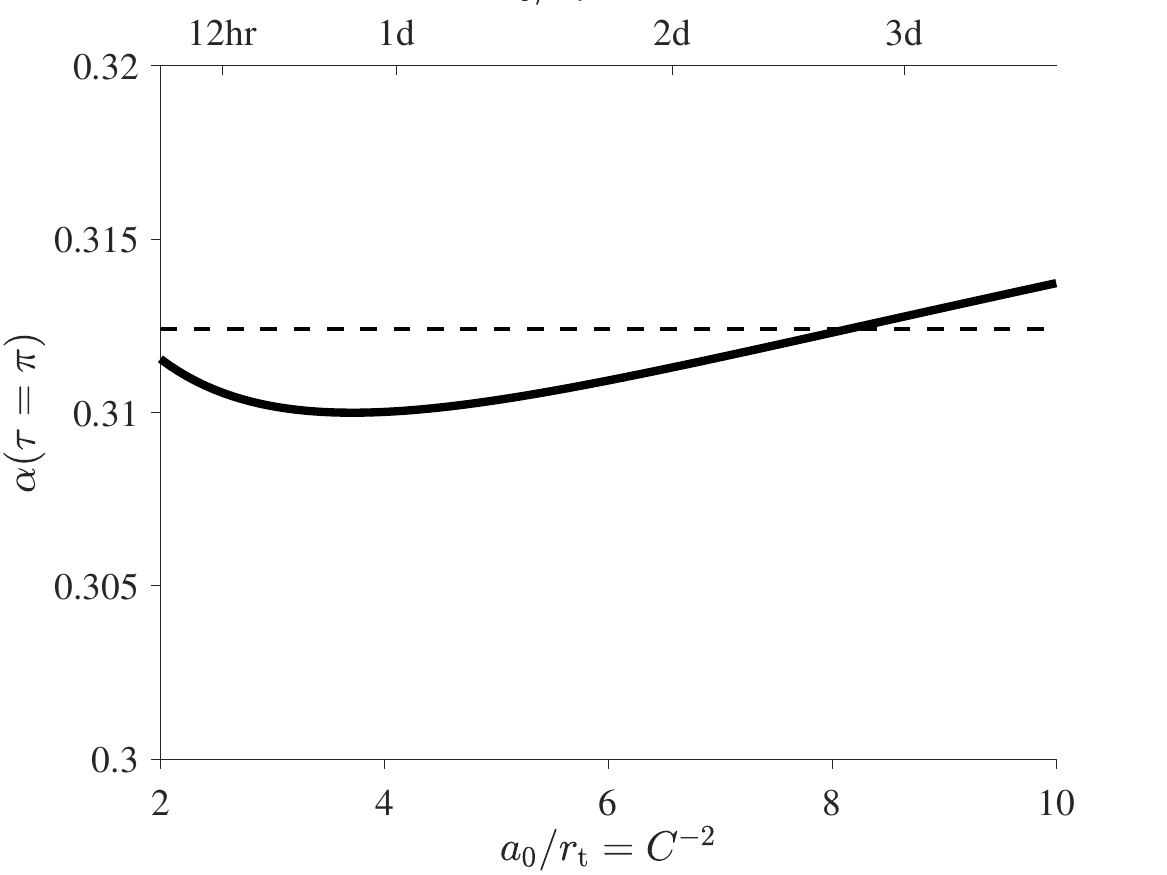}
    \caption{The dimensions and orientation of the stream at time $t=P_{\rm orb}/2$, as a function of $a_0/\rtidal$, or equivalently, $P_{\rm orb}$ (top axis), assuming $m_\star=1 \rm M_\odot$. Top: The stream elongation at $P_{\rm orb}/2$, $\tilde{R}_{1}(\alpha)$. Bottom: The stream angle $\alpha(\pi)$ The dashed horizontal lines correspond to the fiducial values used throughout most of the paper.}
    \label{fig:StreamDimensionsAtCollision}
\end{figure}

\begin{figure}
    \centering
    \includegraphics[width=0.48\textwidth]{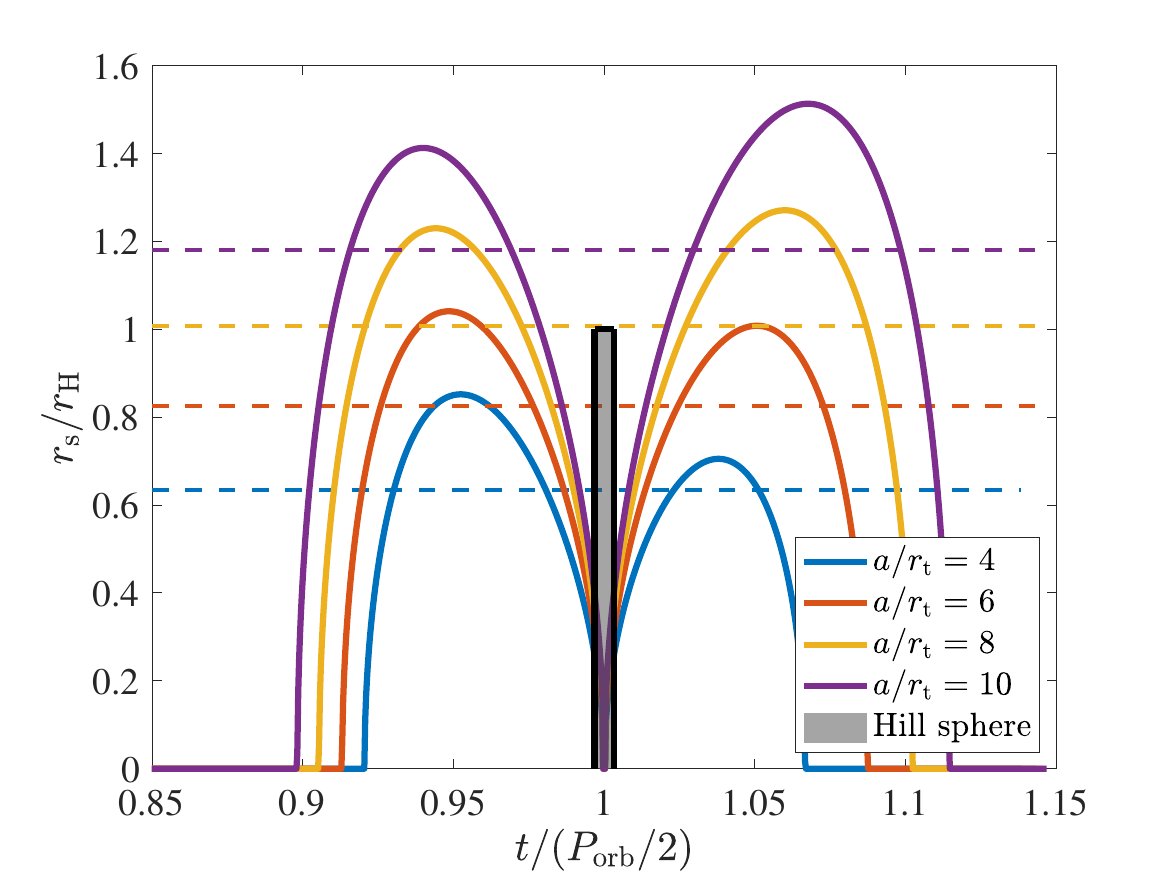}
    \caption{The effective stream width, $r_{\rm s} \approx \sqrt{A_{\rm s}/\pi}$ during the stream-disk encounter. The stream impacts the disk plane with a time-dependent cross section, over the interval $(t_1,t_2)$, $\Delta t_{\rm s}=t_2-t_1$. Maximal compression in the $\tilde{R}_3$ direction (outside the EMRI orbital plane) occurs around $t=P_{\rm orb}/2$, where the effective cross section reaches a minimum. The cross section then increases again after bouncing back, before vanishing at time $t_2$, when the entire stream has impacted the disk. The shaded gray region shows the interval over which the star's Hill sphere is traversing the disk midplane. Solid lines correspond to different values of $a_0/\rtidal$, and horizontal dashed lines show the average value $\sqrt{\left<A_{\rm s}\right> /\pi}$ over the course of encounter.}
    \label{fig:StreamWidth}
\end{figure}

We obtain $\tilde{R}_1(\tilde{t})$, $\tilde{R}_2(\tilde{t})$ by solving $\frac{d}{d\phi}(x^2(\phi;\theta=\frac{\pi}{2}) + y^2(\phi;\theta=\pi/2))=0$ -- i.e., the maximal/minimal displacement from the origin along the $z=0$ plane. The third (out of plane) axis is given by the extent of particles initially located on the poles, ($\theta=0,\pi$)
\begin{equation}
    \tilde{R}_3(\tilde{t}) = \sqrt{1+C^2} \cos{(\tilde{t}-\arctan{C})} \,.
\end{equation}
The general analytical expressions $(\tilde{R}_1(\tilde{t}),\tilde{R}_2(\tilde{t}),\alpha(\tilde{t})$) are not particularly enlightening. The evolution from $t = t_{\rm H}$ to $t \approx P_{\rm orb}/2$, as the stream approaches the disk again, is plotted in Fig.~\ref{fig:FreeSolutionEllipsoid} for $a/\rtidal=5$ (roughly $P_{\rm orb} \approx 30 \,\rm hr$ for a sun-like star). 

\subsection{Stream geometry at subsequent disk crossing}
We are particularly interested in the collision between the debris-stream and the disk around $t\approx P_{\rm orb}/2$ (i.e., $\tilde{t}\approx \pi - 1/C \approx \pi$). As seen in Fig.~\ref{fig:FreeSolutionEllipsoid}, at $\tilde{t}\approx\pi$, the ellipsoidal stream is elongated along the $\tilde{R}_1$ axis. Denoting $\ell \equiv r_{\rm H} \tilde{R}_1(\pi)$ and $\tilde{\ell}\equiv\tilde{R}_1(\pi)$, the stream-disk collision occurs during the time interval $t_1<t<t_2$, with $t_{1,2}\approx P_{\rm orb}/2 \mp \ell /v_{\rm k}$, lasting roughly $\Delta t_{\rm s} = t_2-t_1\approx 2\ell/v_{\rm k}$, neglecting changes in the stream length over the course of the interval $\Delta t_{\rm s}$ (valid for $\Delta t_{\rm s} \ll P_{\rm orb}/2$).

The stream length at $\tilde{t}\approx\pi$ is plotted as a function of $a_0/\rtidal = C^2 \propto P_{\rm orb}^{2/3}$ in Fig.~\ref{fig:StreamDimensionsAtCollision}. Evidently, $\tilde{R}_1(\pi)$ varies by less than a factor of $\sim 2$ over roughly one order of magnitude in orbital period. To facilitate simple analytical estimates, we will use a fiducial, constant value for the stream elongation as it approaches disk impact, $\tilde{\ell} = 35 \, \tilde{\ell}_{35}$.
The stream's inclination upon impact with the disk is nearly independent of $C$, as seen in the bottom panel of Fig.~\ref{fig:StreamDimensionsAtCollision}, forming an angle $\alpha(\pi)\approx 0.31 \, {\rm rad} \simeq 18^{ \circ}$ with the normal to the disk plane.

In addition to tidal stretching in the orbital plane, the ejecta is compressed to a thin elliptical ``pancake'', with $\tilde{R}_3(t_{\rm c})=0$ at time 
\begin{equation} \label{eq:tau_max_compression}
    \tilde{t}_{\rm c} = \pi-\arctan{(1/C)} \,,
\end{equation}
near the star's subsequent disk passage \footnote{Maximal compression reflects the intersection of inclined ejecta orbits as they cross the line of nodes, akin to the pericenter "nozzle shock" compression occurring in TDE debris streams  \citep[e.g.,][]{Carter_Luminet_1983,Stone_2013,Bonnerot_Lu_22}. While the compression in the highly-eccentric debris stream of a TDE is localized to a point, here, the entire ellipsoid is compressed to a pancake at (dimensionless) time $\tilde{t}_{\rm c}$.}. We stress that for homologous expansion (i,e., uniform $C$ for all shells), maximal vertical compression occurs simultaneously for all concentric ellipsoids. The simultaneity of fluid compression is expected to disappear in the presence of orbital eccentricity, perturbations to the free solution (e.g., star's gravity, stream self-gravity, pressure gradients). If, relative to the compression timescale, $|z/\dot{z}|$, the compressed stream cools slowly, the build-up of pressure opposes the vertical collapse until it stalls and bounces back, inflating back along a trajectory which traces the time reversal of the motion in the $z$ direction (intuitively, as if collapsing stream would self-cross, unperturbed). In the adiabatic limit (i.e., $p\propto \rho^\gamma$) maximal compression and reversal of the vertical collapse occur roughly when the stream's overall volume is shrunk down to its initial volume when it was in rough virial equilibrium, i.e., $\sim R_\star^3$. This therefore occurs as $\tilde{R_1}\tilde{R}_2\tilde{R_3}\lesssim (\rtidal/a_0)^3 = C^{-6}$. We will explore additional aspects of the stream vertical compression and its potential observational implications in future work.

Accounting for the evolving ellipsoid dimensions and its orientation, as well as the motion of the disk plane (relative to the star's frame), we obtain the stream-disk area of intersection, along the collision time interval $t_1<t<t_2$. We express the momentary intersection area as $A_{\rm s} \equiv \pi r_{\rm s}^2$, where $r_{\rm s}$ is the geometric mean of the axes of the elliptical section (trivially, the intersection of any plane crossing through the ellipsoidal stream forms an ellipse). Fig.~\ref{fig:StreamWidth} shows $r_{\rm s}$ (in units of $r_{\rm H}$) as a function of time. At $t>t_1$, $r_{\rm s}$ initially grows as wider parts of the ellipsoid are traversing the disk plane, until vertical compression begins to dominate, decreasing the intersection area and $r_{\rm s}$ towards a minimum at $t=t_{\rm c}$. The increasing vertical pressure gradient causes the stream to reverse its collapse and bounce back, increasing the collision cross section, until the entire stream has traversed the disk at $t=t_2$. 
Evidently, for a range of $a/\rtidal$, the mean cross section is approximately $\left< A_{\rm s} \right> \approx \pi r_{\rm H}^2$. Furthermore, the time of maximal vertical compression is also when the star (located at the center of the stream) passes through the disk. As the star itself is engulfed by the marginally bound ejecta remaining in its Hill sphere, the collision cross section is roughly $\approx \pi r_{\rm H}^2$ for a duration $\Delta t_{\rm H} \approx r_{\rm H}/v_{\rm k} \approx (m_\star/\MBH)^{1/3} \Omega_0^{-1}$ around $\tilde{t}_{\rm c}$, such that $r_{\rm s} \to 0$ is never realized, even in this idealized picture (Fig.~\ref{fig:StreamWidth}).

In summary, ablation and tidal evolution of the stellar debris result in an elongated stream of length $2\ell \approx 70 \, r_{\rm H} \, \tilde{\ell}_{35}$ and a (time averaged) transverse cross section $A_{\rm s} \approx \pi r_{\rm H}^2$, with mean stream density of $\rho_{\rm s} \approx m_{\rm ej,\star}/(70\pi \tilde{\ell}_{35} r_{\rm H}^3)$, colliding with the disk over a timescale $\Delta t_{\rm s} \approx 2\ell /v_{\rm k} \approx 0.1 \, P_{\rm orb}$.

\subsection{Debris-stream geometry and impacted disk}
\label{sec:EjectaStreamCollisionWithDisk}

The geometry of the stream relative to the disk (Fig.~\ref{fig:SchematicOverview}) implies that its footprint on the surface of the disk stretches along a radial range of approximately $\delta r/a_0 \approx 2\tilde{\ell} \, \sin{\alpha} (m_\star/\MBH)^{1/3} \approx 0.2$, and an azimuthal range $\delta \varphi\approx 2\tilde{\ell} (m_\star/\MBH)^{1/3} \cos{\beta} \approx 0.7 \, \rm rad \approx 40^\circ$, traversed during the disk crossing time. The disk mass participating in the stream impact is roughly 
\begin{multline} \label{eq:m_shocked_disk_mass}
    m_{\rm s,d} \approx 4 \, \Sigmad r_{\rm H}^2\tilde{\ell} \approx \\
    2\times10^{-5} \, {\rm M_\odot} \; \rho_{\rm d,-7} \tilde{h}_{-2} m_1^{2/3} P_{\rm orb,8}^2 M_{\bullet,6}^{1/3} \tilde{\ell}_{35} \,.
\end{multline}
Much of the subsequent evolution depends on the ratio $m_{\rm ej,\star}/m_{\rm s,d}$
\begin{equation} \label{eq:m_ej_vs_m_disk}
    \frac{m_{\rm ej,\star}}{m_{\rm s,d}} \approx 1 \,
    P_{\rm orb,8}^{-8/3}\;\eta_{0.03} \, M_{\bullet,6}^{1/3} m_1^{1.5}  (\tilde{h}_{-2} \tilde{\ell}_{35})^{-1} \,.
\end{equation}
In the limit $m_{\rm ej,\star}\ll m_{\rm s,d}$, the ejecta stream is effectively decelerated and deflected as it collides with the disk, with most of the mass $m_{\rm s,d}$ remaining mostly unperturbed. In the complementary limit, $m_{\rm ej,\star} \gg m_{\rm s,d}$, the ejecta stream cuts through the disk, with the mass $m_{\rm s,d}$ being strongly shocked by the incoming stream which acts as a barrier blocking the disk's local azimuthal flow. In this regime, it is the shocked mass $m_{\rm s,d}$ which will dominate the emission, as we later discuss. The critical orbital period delineating these two regimes is roughly 
\begin{equation} \label{eq:P_mej=msd}
    P_{\rm orb}(m_{\rm ej,\star}=m_{\rm s,d}) \approx 10 \, {\rm hr} \; m_1^{0.56} M_{\bullet,6}^{1/8} \left( \tilde{h}_{-2} \tilde{\ell}_{35} /\eta_{0.03} \right)^{-3/8} \,.
\end{equation}

The stream mass, $m_{\rm ej,\star}$ is distributed over a volume ${\sim}2\pi \tilde{\ell} r_{\rm H}^3/3$ at time $t\approx P_{\rm orb}/2$, such that the stream density relative to that of the disk midplane is
\begin{multline} \label{eq:rho_s_relative}
    \frac{\bar{\rho}_{\rm ej,\star}}{\rhod} \approx \frac{\bar{\rho}_{\rm ej,i}}{\rhod} \tilde{\ell}^{-1}
    \approx 0.8 \; \eta_{0.03} M_{\bullet,6}^{2/3} m_1^{1.2} P_{\rm orb,8}^{-8/3} \tilde{\ell}_{35}^{-1} \,,
\end{multline}
The stream's mass current is approximately
\begin{multline}
    \dot{m}_{\rm s,ej} \approx \frac{m_{\rm ej,\star}}{\Delta t_{\rm s}} \approx \pi \bar{\rho}_{\rm ej,\star} r_{\rm H}^2 v_{\rm k} \approx \pi \eta \rhod a_0^2 v_{\rm k} \pfrac{\rtidal}{a_0}^4 \tilde{\ell}^{-1} \\
    \approx 6 \, \dotMEdd \; \tilde{\ell}_{35}^{-1} \eta_{0.03} \,\rho_{\rm d,-7} \, M_{\bullet,6}^{-1/3} m_1^{2.2} P_{\rm orb,8}^{-5/3} \,,
\end{multline}
where the subscript $\rm ``s,ej"$ refers to the stream of stellar ejecta impacting the disk. 

For comparison, the current carried by the impacted disk annulus is set by the projected width of the impacting stream in the radial direction
\begin{multline}
    \dot{m}_{\rm s,d} \approx 
    2 r_{\rm H} \Sigmad v_{\rm k} \approx \\
    7 \, \dotMEdd \; \rho_{\rm d,-7} \tilde{h}_{-2} m_1^{1/3} M_{\bullet,6}^{-2/3} P_{\rm orb,8} \,,
\end{multline}
where the subscript $\rm ``s,d"$ refers to the ``stream" of disk material colliding with the ejecta stream.

\begin{figure}
    \centering
    \includegraphics[width=0.49\textwidth]{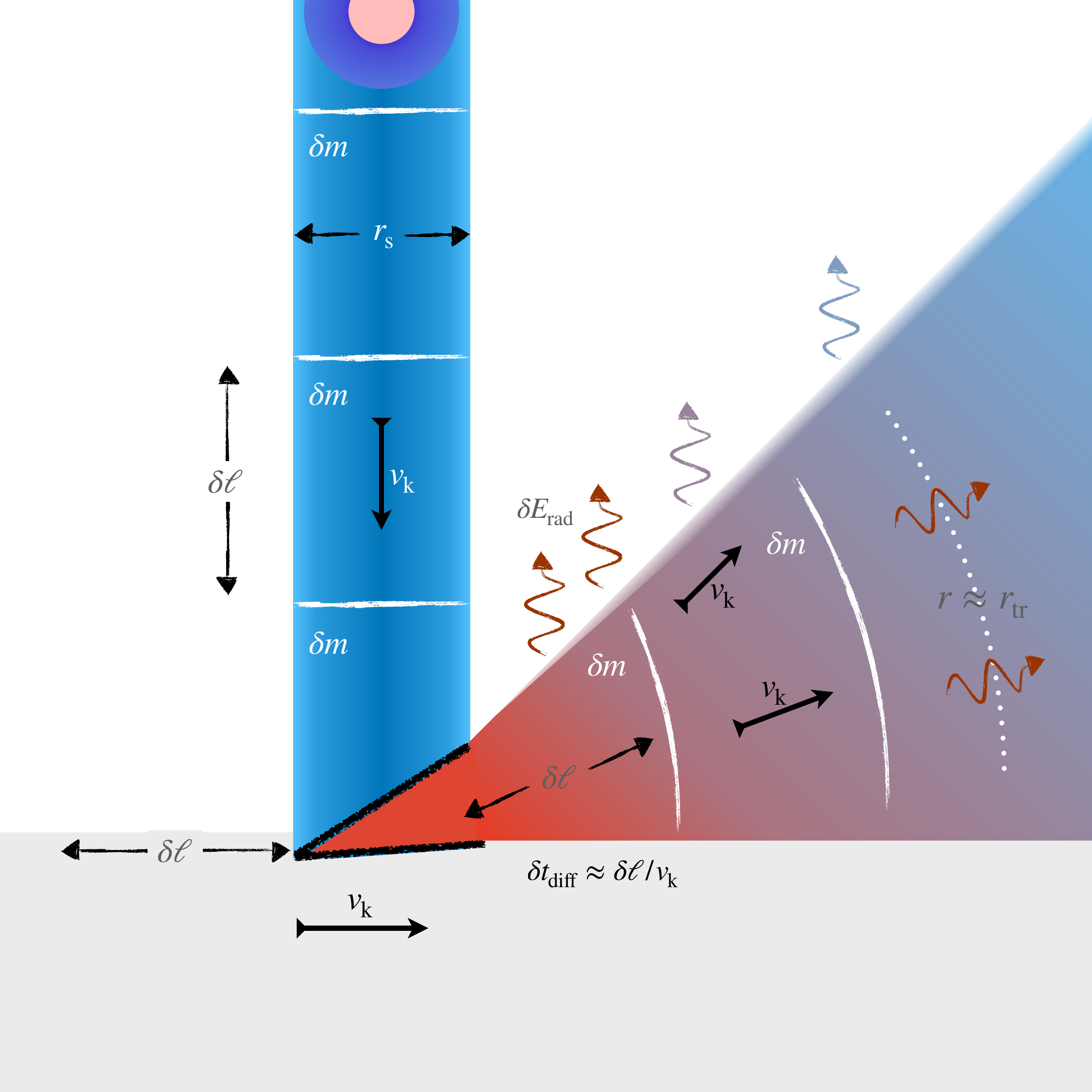}

    \caption{Schematic sketch of the stream-disk interaction. An elongated stream of stellar material impacts the disk, which is rotating azimuthally (to the right in the image). A strong reverse shock forms at the disk/stream interface, where the shocked stream is heated and deflected to form a wedge of quasi-spherical outflow. Radiation either escapes promptly near the stream base, or it is advected along with the optically thick outflow up to the trapping radius (dotted white line on the right).}
    \label{fig:Cartoon_Star_Stream_Disk_Interaction}
\end{figure}

\section{Observable Signatures of Star-Disk and Stream-Disk Collisions} \label{sec:Observables}

Equipped with the overall properties of the ablated stellar-debris stream upon its impact with the disk, we now focus on observable signatures of the different components of the star-disk collision. We begin by briefly reciting the analytical estimates of \cite{Linial_Metzger_23} regarding the emission that follows the collision of a (bare) star with a disk. We then account for the increased effective cross section of the star, in the presence of dense ablated stellar material within its Hill sphere. The key new results presented here address the interaction between the disk and the stream of ablated stellar debris, and its observational signatures.

Throughout this section, we remain fairly agnostic to the exact disk properties (i.e., $\rhod$ and $\tilde{h}$), and scale by arbitrary fiducial values. More concrete assumptions about the underlying disk are discussed in \S \ref{sec:DiskModels}. 

While the analysis presented here provides a general picture of the emission timescales and energetics, it is not intended to substitute a more careful analysis of the relevant radiative processes. For example, the stream-disk collision studied in \cite[e.g.,][]{Chan_2021_tde_agn} highlights possibilities not addressed here, such as inverse-Compton cooling of shock heated ejecta, and the high-energy photons it may produce. Treatment of the emission in photon starved ejecta, as those invoked in \cite{Linial_Metzger_23}, and in more detail in \cite{Vurm_25} are also neglected here. Particularly, the emission temperatures inferred in this section typically represent a lower limit on the effective photon energies. This limit is valid if photon production and thermalization occur rapidly, such that blackbody thermal equilibrium can be assumed for the escaping radiation.

\subsection{Bare Star \& Disk} \label{sec:bare_star+disk}
Considering the impact of a bare star (i.e., without an extended ablated debris), the shock-heated disk mass is roughly $m_{\rm ej,d}\approx 2\pi R_\star^2 \Sigmad$. It initially possesses approximately equal amounts of kinetic and internal energy, $E_i \approx m_{\rm ej,d} v_{\rm sh}^2 / 2$, where $v_{\rm sh} \approx v_{\rm ej} \approx v_{\rm k}\sqrt2$ is the shock/ejecta velocity. The internal radiation energy is trapped in the initially optically thick ejecta, escaping as the optical depth drops to roughly $\tau_{\rm rad} \approx c/v_{\rm ej}$. During the quasi-spherical expansion phase of the ejecta, the optical depth decreases with time as $\tau(t) \approx \kappa m_{\rm ej,d}/(4\pi (v_{\rm ej} t)^2) \approx \kappa \Sigmad (t/t_i)^{-2}$
where $t_i \approx V_i^{1/3} /v_{\rm k}$ is the initial expansion timescale of the shock heated disk material, of volume $V_i\approx 2\pi R_\star^2 \Hd/7$, and the factor $1/7$ accounts for the strong-shock compression of a radiation-pressure dominated medium. The photon diffusion time, $t_{\rm diff}$, defined by $\tau(t_{\rm diff,d})\approx \tau_{\rm rad}$, is roughly
\begin{multline} \label{eq:t_diff_star+d}
    t_{\rm diff,\star+d} \approx \sqrt{\frac{\kappa m_{\rm ej,d}}{4\pi c v_{\rm k}}}
    \approx 6 \, {\rm min} \, R_1 \rho_{\rm d,-7}^{1/2} \tilde{h}_{-2}^{1/2} P_{\rm orb,8}^{1/2} \,,
\end{multline}
where $R_1 = R_\star/{\rm R_\odot}$. During its adiabatic expansion, the ejecta's internal/radiation energy is reduced due to $PdV$ work, with the bulk of the remaining energy radiated over a timescale $t_{\rm diff,d}$, with $E_{\rm rad} \approx E_i (t_i/t_{\rm diff,d})$, corresponding to the adiabatic index $\gamma_{\rm ad}=4/3$ of a radiation-pressure dominated gas. The resulting bolometric luminosity can be approximated as
\begin{multline} \label{eq:L_bol_star+d}
    L_{\rm \star+d} \approx \frac{E_{
    \rm rad}}{t_{\rm diff}} \approx \frac{E{\rm i}}{t_{\rm diff}^2} t_i \approx L_{\rm Edd} \frac{(R_\star^2 \Hd)^{1/3}}{a_0} \approx \\
    10^{42} \, {\rm erg \, s^{-1}} \; M_{\bullet,6}^{7/9} P_{\rm orb,8}^{-4/9} R_1^{2/3} \tilde{h}_{-2}^{1/3}\,,
\end{multline}
notably independent of $m_{\rm ej,d}$, and only weakly dependent on the disk aspect ratio, $\tilde{h}$. The total radiated energy, $L_{\star+\rm d} t_{\rm diff}$ is then approximately
\begin{multline} \label{eq:E_rad_star+disk}
    E_{\rm rad}
    \approx 4\times 10^{44} \, {\rm erg} \;
    R_1^{5/3} \rho_{\rm d,-7}^{1/2} \tilde{h}_{-2}^{5/6} P_{\rm orb,8}^{1/18} M_{\bullet,6}^{7/9} \,.
\end{multline}

The effective blackbody temperature of the bulk of the emission is approximately
\begin{multline}
    \kb T_{\rm BB} \approx \kb \pfrac{E_{\rm rad}}{4\pi (v_{\rm k} t_{\rm diff})^3a_{\rm rad}}^{1/4} \approx \\
    20 \, {\rm eV} \; R_1^{-1/3} \rho_{\rm d,-7}^{-1/4} \tilde{h}_{-2}^{-1/6} P_{\rm orb,8}^{-1/9} M_{\bullet,6}^{-1/18} \,,
\end{multline}
where we assumed rapid photon production and thermalization in the shocked medium. However, photon starvation effects \citep{Linial_Metzger_23,Vurm_25} may increase the effective observed temperature, in particular, at low disk densities and high collision velocities (e.g., Figure 16 of \citealt{Vurm_25}).

\subsection{Star+Hill Sphere \& Disk} \label{sec:Hill+disk}
A fraction of the ablated stellar material remains marginally bound to the star, lingering within its Hill sphere before the next disk collision. The combined structure of the star, engulfed with an extended envelope of ablated material, subtends a radius $r_{\rm H}$, which acts as an effectively larger ``star''. Substituting $R_\star=r_{\rm H}$ in Eq.~\ref{eq:t_diff_star+d}, we find the diffusion time of the shocked disk material, $m_{\rm d,H}$ (Eq.~\ref{eq:m_disk_Hill})
\begin{equation} \label{eq:t_diff_H+d}
    t_{\rm diff,H+d} \approx 12 \, {\rm min} \; P_{\rm orb,8}^{7/6} m_1^{1/3} ( \rho_{\rm d,-7} \tilde{h}_{-2} )^{1/2} \,,
\end{equation}
and the luminosity can be similarly obtained from Eq.~\ref{eq:L_bol_star+d}
\begin{multline}
    L_{\rm H+d} \approx L_{\rm Edd} (r_{\rm H}^2 \Hd)^{1/3}/a_0 \approx \\
    1.5\times10^{42} \, {\rm erg \, s^{-1}} \; M_{\bullet,6}^{7/9} m_1^{2/9} \tilde{h}_{-2}^{1/3}\,,
\end{multline}
independent of orbital period, and only weakly dependent on $m_1$ and $\tilde{h}$. The total radiated energy is
\begin{multline} \label{eq:E_rad_H+disk}
    E_{\rm rad, H+d} \approx 
    10^{45} \, {\rm erg} \; M_{\bullet,6}^{7/9} m_1^{5/9} \tilde{h}_{-2}^{5/6} P_{\rm orb,8}^{7/6} \rho_{\rm d,-7}^{1/2} \,,
\end{multline}
and the blackbody temperature,
\begin{equation}
    \kb T_{\rm BB} \approx 15 \, {\rm eV} \; m_1^{-1/9} \rho_{\rm d,-7}^{-1/4} \tilde{h}_{-2}^{-1/6} P_{\rm orb,8}^{-1/3}  M_{\bullet,6}^{-1/18} \,,
\end{equation}
with similar caveats regarding photon starvation discussed above which may lead to higher effective temperatures.

We note that the enhancement in the star's effective size to $\sim r_{\rm H}$ requires that the (marginally bound) ejecta mass in the Hill sphere exceeds the impacted disk mass, $m_{\rm d,H} \approx\pi r_{\rm H}^2 \Sigmad$. Since the remaining ejecta mass within $r_{\rm H}$ is of the order of $m_{\rm ej,\star}$ (comparable mass in bound and unbound debris, \citealt{Yao_2025}), this criterion is equivalent to $\tau_{\rm ej,i} \gtrsim \kappa_{\rm es} \Sigmad$ (Eq.~\ref{eq:optical_depth_inital}), or
\begin{equation}
    \frac{\tau_{\rm ej,i}}{\kappa_{\rm es} \Sigmad} \approx \frac{m_{\rm ej,\star}}{\pi r_{\rm H}^2 \Sigmad} \approx 30 \; \eta_{0.03} M_{\bullet,6}^{2/3} m_1^{1.5} P_{\rm orb,8}^{-8/3} \tilde{h}_{-2}^{-1} \,.
\end{equation}

The regime where the disk is shocked by dense stellar debris enclosed in the star's Hill sphere is therefore valid only up to a maximal orbital period
\begin{equation}
    P_{\rm orb}^{\rm max, H+d} \approx 1.3 \, {\rm d} \; \eta_{0.03}^{3/8} M_{\bullet,6}^{1/4} m_1^{0.6} \tilde{h}_{-2}^{-3/8} \,,
\end{equation}
such that at yet longer orbital periods, it is only the ``bare'' star that serves as the effective impactor shocking disk material, reverting to the emission discussed in \S \ref{sec:bare_star+disk}. In the limit $m_{\rm H,d} \gg m_{\rm ej,\star}$, the ejecta within the Hill sphere is decelerated and shocked by the more massive disk, effectively contributing to the shocked debris stream, as discussed in the following section.

\begin{figure}
    \centering
    \includegraphics[width=1.1\linewidth]{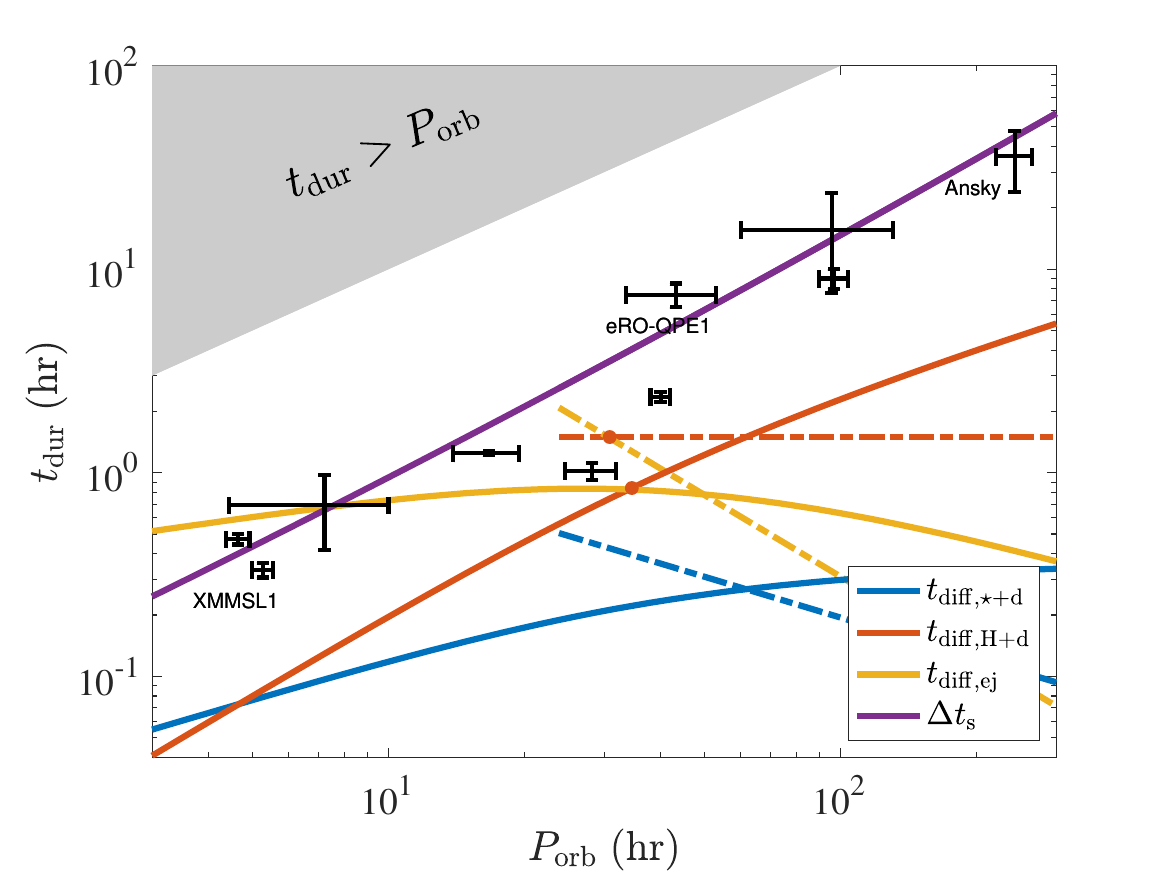}
    \caption{Different emission timescales versus the EMRI orbital period. Plotted are the photon diffusion timescales of the shocked disk material, when the effective cross section is set by the size of star (\textit{blue}) or its Hill sphere (\textit{red}), and the photon diffusion time of the shock-heated stellar ejecta, in the explosive regime (\textit{yellow}). The underlying disk is either a steady Shakura-Sunyaev accretion disk with $\dot{m}_{\rm d}=0.1\, M_{\rm Edd}$, $\alpha=0.1$, $\kappa=\kappa_{\rm es}$ (\textit{solid lines}), or, the outskirts of a viscously spreading TDE disk with conserved angular momentum, for $m_{\star}^{\rm tde}=1 \, \rm M_\odot$ (\textit{dash-dotted lines}, limited to $P_{\rm orb} \gtrsim 1 \, \rm d$, see \S \ref{sec:DiskModels}). Lastly, the collision duration of a tidally stretched stream of ablated stellar material is shown in \textit{purple}. The black error bars indicate the range of timescales observed for various QPE sources, where $P_{\rm orb} = 2 \left< P_{\rm QPE}\right>$ was assumed.}
    \label{fig:TimescaleComparison}
\end{figure}

\subsection{Debris Stream \& Disk}

Upon impacting the disk, the stream's kinetic energy is converted to internal energy through a strong reverse shock which redirects and thermalizes the stream (Fig.~\ref{fig:Cartoon_Star_Stream_Disk_Interaction})\footnote{The scenario studied here is similar to that of \cite{Chan_2021_tde_agn} in the context of TDEs in AGN, and the second encounter between the TDE bound debris and the AGN disk (see their Figure 5). We discuss this point further in \S \ref{sec:Discussion}.}. In the ``explosive'' regime, the collision timescale is short compared to the shocked ejecta's overall cooling time, and the stream's total kinetic energy $E_{\rm i} \approx m_{\rm ej,\star} v_{\rm k}^2 / 2$ is deposited as a ``thermal bomb'' around the stream-disk collision site. As in the previous calculations, most of the remaining internal energy is radiated when $\tau_{\rm rad}=c/v_{\rm k}$, at time
\begin{multline}
    t_{\rm diff,tot} \approx \sqrt{\frac{\kappa m_{\rm ej,\star}}{4\pi c v_{\rm k}}} \\
    \approx 50 \, {\rm min} \; \rho_{\rm d,-7}^{1/2} \eta_{0.03}^{1/2} (M_{\bullet,6}/P_{\rm orb,8})^{1/6} m_1^{1.1} \,,
\end{multline}
where here the shocked mass dominating the emission is $m_{\rm ej,\star}$, rather than shocked disk material ($m_{\rm ej,d}$ or $m_{\rm d,H}$). Compared to the stream-disk collision timescale
\begin{multline} \label{eq:t_diff_tot_verus_ts}
    \frac{t_{\rm diff,tot}}{\Delta t_{\rm s}} \approx \sqrt{\tau_{\rm ej,i} \frac{v_{\rm k}}{c}} \frac{1}{2\tilde{\ell}} \\
    \approx 0.9 \; ( \eta_{0.03} \, M_{\bullet,6} \rho_{\rm d,-7})^{1/2} \tilde{\ell}_{35}^{-1} m_1^{0.75} P_{\rm orb,8}^{-7/6} \,,
\end{multline}
such that at sufficiently short periods ($P_{\rm orb} \lesssim 8 \, {\rm hr} \; m_1^{0.64} \, \rho_{\rm d,-7}^{3/7}$, where other parameters are fixed), the collision between the stream and the disk is effectively ``explosive'' - radiation is trapped long after the entire stream has impacted the disk. If $m_{\rm ej,\star} \ll m_{\rm s,d}$ (see Eq.~\ref{eq:m_ej_vs_m_disk}), it is the shocked ejecta governing the emission, radiating with luminosity
\begin{multline}
    L_{\rm s+d} \approx L_{\rm Edd} (r_{\rm H}/a_0)\tilde{\ell}^{1/2} \approx \\
    9 \times 10^{42} \, {\rm erg \,s^{-1}} \; (m_1/M_{\bullet,6})^{1/3} \tilde{\ell}_{35}^{1/2} \,,
\end{multline}
where we assumed that the shocked stream expands and cools from an initial length scale $\sim \tilde{\ell}^{1/2} r_{\rm H}$ (dictated by the area of the collision region on the disk, of length $2\ell$ and width $r_{\rm H}$). However, since energy is deposited along an elongated arc of rather than a localized collision site (as in the case of a star impacting the disk), our previous arguments are not directly applicable in this scenario.

On the other hand, for sufficiently long orbital periods ($P_{\rm orb} \gtrsim 8 \, \rm hr$ for the same fiducial parameters), the flare duration is set by the \textit{stream arrival timescale}
\begin{equation} \label{eq:t_stream_duration}
    \Delta t_{\rm s} \approx 0.1 \, P_{\rm orb} \; m_1^{1/3} M_{\bullet,6}^{-1/3} \tilde{\ell}_{35} \,,
\end{equation}
rather than by the photon diffusion time associated with the shocked ejecta. 

Fig.~\ref{fig:TimescaleComparison} compares the various timescales associated with the different emission components (diffusion times, stream collision duration), as a function of $P_{\rm orb}$, alongside the durations and recurrence intervals of 10 QPE sources. We assume here two possibilities for the underlying disk: A spreading TDE disk and a steady-state alpha-disk (discussed in more detail in \S \ref{sec:DiskModels}). It is evident that the stream arrival time, $\Delta t_{\rm s}$ matches the observation, while photon diffusion timescales generally fail to explain flare durations of long $(P_{\rm orb} \gtrsim 1 \, \rm d)$ QPEs. We stress that the purple curve of Fig.~\ref{fig:TimescaleComparison} is not a fit to the data, but rather calculated directly from the stream geometry given by our analytical derivations of \S \ref{sec:EjectaStreamCollisionWithDisk}.

We now turn to discuss the emission arising from the prolonged interaction between a tidally elongated debris stream and the disk, in the limit $t_{\rm diff,tot}\ll \Delta t_{\rm s}$. We consider two idealized limits of the resulting flow and emission that follow the collision:
\begin{itemize}
    \item \textit{Wind-like emission}: Where the impacting stream produces a wind-like optically thick outflow enshrouding the collision site, with the emission launched from an extended photon trapping radius set by the stream's mass content.
    \item \textit{Sequence of explosions}: Where the stream-disk collision is described as a continuum of disjoint ``explosions'', each involving a small fraction of the total stream mass.
\end{itemize}

\subsubsection{Shocked Stream: Wind-like Outflow and Emission} \label{sec:Stream_Wind}

An elongated stream of ablated stellar material of total mass $m_{\rm ej,\star}$, length $2\ell$, typical cross section and mass density $A_{\rm s} = \pi r_{\rm s}^2$ and $\rho_{\rm ej,s}$ impacts the rotating disk. As argued before, for sufficiently long orbital periods, the ejecta mass is small relative to the disk mass participating in the collision, $m_{\rm ej,\star} \ll m_{\rm s,d}$ (see Eq.~\ref{eq:m_ej_vs_m_disk} and accompanying discussion), implying that the stream is strongly shocked and deflected by the denser accretion flow. Here we assume that the resulting outflow has a constant velocity $v_{\rm w} \approx v_{\rm s}$, launched from the impact site, producing a wind-like density profile of shocked stream material
\begin{equation}
    \rho_{\rm w}(r) \approx \frac{\dot{m}_{\rm s}}{\Delta \Omega \, r^2 \, v_{\rm w}} \,,
\end{equation}
where $\Delta \Omega$ is the solid angle subtended by the outflow. We anticipate $\Delta \Omega\sim 1$, as the outflow is limited to one side of the disk (in the limit $m_{\rm ej,\star}\ll m_{\rm s,d}$), and it is expected to be mostly lateral due to the disk rotation. The typical length scale connecting the linear, unshocked stream and the quasi-spherical outflow is $r_{\rm s} \equiv \sqrt{A_{\rm s}/\pi} \approx r_{\rm H}$, set by the dimensions of the impacting stream. Assuming constant opacity, the outflow's optical depth from radius $r$ to infinity is $\tau_{\rm w}(r)=\int_r^\infty \kappa \rho_{\rm w}(r') \, {\rm d}r' \approx \kappa \rho_{\rm w}(r)r$. Neglecting photon diffusion in the azimuthal direction or from the rims of the outflow (see Fig.~\ref{fig:Cartoon_Star_Stream_Disk_Interaction}), radiation is advected along with the wind from the inner radius $r_{\rm s}$ up to the \textit{trapping radius}, $r_{\rm tr}$, where $\tau_{\rm w}(r_{\rm tr})\approx c/v_{\rm w}$, $r_{\rm tr} \approx \kappa \dot{m}_{\rm s}/(\Delta\Omega\, c)$. Relative to the outflow launching scale, $r_{\rm s}$
\begin{multline}
    \frac{r_{\rm tr}}{r_{\rm s}} \approx \frac{\pi}{\Delta\Omega} \pfrac{v_{\rm k}}{c} \tau_{\rm s,\perp} \approx
    370 \; \frac{\eta_{0.03} \,\rho_{\rm d,-7}}{(\Delta \Omega/\pi)} \\ \tilde{\ell}_{35}^{-1}
    P_{\rm orb,8}^{-7/3} m_1^{1.5} M_{\bullet,6}^{2/3} \,.
\end{multline}
where $\tau_{\rm s,\perp} \approx\kappa \rho_{\rm s} r_{\rm s}$ is the stream's optical depth in the direction perpendicular to its propagation. We will later focus on the limit $r_{\rm tr}/r_{\rm s} \approx 1$ is where the emission behind the shock can escape promptly -- we discuss this point in the next section (\S \ref{sec:Efficiency}).

A quasi-steady flow is established up to the trapping radius over a timescale $t_{\rm tr} \approx r_{\rm tr}/v_{\rm w}$. Compared to the duration of stream-disk collision
\begin{multline} \label{eq:timeratio_rhod_dom}
    \frac{t_{\rm tr}}{\Delta t_{\rm s}} \approx 
    \frac{r_{\rm tr}}{2\ell} = 
    \frac{\kappa \dot{m}_{\rm s}}{2\Delta \Omega c r_{\rm H}} \tilde{\ell}^{-1} \approx \\
    \approx 2.5 \; \frac{\eta_{0.03} \,\rho_{\rm d,-7}}{(\Delta \Omega/\pi)} \tilde{\ell}_{35}^{-2} M_{\bullet,6}^{1/3} m_1^{2.2} P_{\rm orb,8}^{-7/3}  \,, 
\end{multline}
and thus the wind approximation is valid for $\rho_{\rm d,-7} P_{\rm orb,8}^{-7/3} \lesssim 0.5$ (with the other parameters fixed). Otherwise, the entire stream is shocked before the outflow reaches scales of order $r_{\rm tr}$, reverting to the instantaneous energy injection limit described above.

When $t_{\rm tr} \ll \Delta t_{\rm s}$, the emission is sourced from radius $r_{\rm tr}$ within the quasi-steady wind. Since the flow evolves adiabatically in the range $r_{\rm s} \lesssim r \lesssim r_{\rm tr}$, the radiation energy density scales as $u_{\rm rad} \propto \rho_{\rm w}^{\gamma} \propto r^{-8/3}$ (with $\gamma=4/3$ for a radiation pressure dominated gas). On the inner scale, $r_{\rm s}$, where the stream is shocked and the outflow is launched, $u_{\rm rad}(r_{\rm s})\approx\rho_{\rm s} v_{\rm s}^2$ (assuming that the downstream is radiation dominated). Thus, the resulting wind luminosity is
\begin{multline} \label{eq:L_bol_stream_disk_wind}
    L \approx 
    \Delta \Omega \, r_{\rm tr}^2 u_{\rm rad}(r_{\rm tr}) c / \tau_{\rm w}(r_{\rm tr}) \approx 
    \Delta \Omega \; \dot{m}_{\rm s}v_{\rm s}^2 \pfrac{r_{\rm tr}}{r_{\rm s}
    }^{-2/3} \\
    \approx 10^{43} \, {\rm erg \, s^{-1}} \;
    (\Delta \Omega/\pi)^{5/3} (\eta_{0.03} \rho_{\rm d,-7} /\tilde{\ell}_{35})^{1/3} \\
     m_1^{0.96} P_{\rm orb,8}^{-7/9} M_{\bullet,6}^{8/9} \,.
\end{multline}
The total radiated energy
\begin{multline}
    E_{\rm rad} \approx L \Delta t_{\rm s} \approx
    3\times 10^{46} \, {\rm erg} \; (\Delta \Omega/\pi)^{5/3}  P_{\rm orb,8}^{2/9} M_{\bullet,6}^{5/9} \\
    (\eta_{0.03} \rho_{\rm d,-7}/\tilde{\ell}_{35})^{1/3} m_1^{1.3}  \,,
\end{multline}
and the radiative efficiency is roughly
\begin{multline} \label{eq:efficiency_wind_params}
    \varepsilon_{\rm rad} \approx \frac{E_{\rm rad}}{m_{\rm ej,\star} v_{\rm k}^2} \approx 0.07 \; (\Delta \Omega/\pi)^{5/3} \tilde{\ell}_{35}^{-1/3} \\
    \eta_{0.01}^{-2/3} \rho_{\rm d,-7}^{-2/3} m_1^{-0.9} P_{\rm orb,8}^{14/9} M_{\bullet,6}^{-7/9} \,.
\end{multline}

As the remaining optical depth above $r_{\rm tr}$ and up to $\tau\approx 1$ is scattering dominated, both the luminosity and observed temperature are set at $\tau_{\rm w}(r_{\rm tr})\approx c/v_{\rm w}$. Assuming efficient thermalization, the blackbody temperature is approximately
\begin{multline} \label{eq:T_BB_wind}
    \kb T_{\rm BB} \approx \kb \pfrac{u_{\rm rad}}{a_{\rm rad}}^{1/4} \approx \kb \pfrac{\rho_{\rm s} v_{\rm k}^2}{a_{\rm rad}}^{1/4} \pfrac{r_{\rm tr}}{r_{\rm s}}^{-2/3} \approx \\
    7 \, {\rm eV} \,(\tilde{\ell}_{35}/\eta_{0.03}\rho_{\rm d,-7})^{5/12} P_{\rm orb,8}^{13/18} (\Delta \Omega/\pi M_{
    \bullet,6})^{2/3} m_1^{-0.7}
\end{multline}
and the temperature grows with increasing period and decreasing disk density, $T_{\rm BB} \propto \rhod^{-5/12} P_{\rm orb}^{13/18}$. This primarily reflects the reduced ejecta mass, resulting in lower optical depth, and flare temperatures that approach the postshock temperature.

\subsubsection{Stream collision as a sequence of explosions} \label{sec:stream_sequence_of_collisions}
The impacted disk annulus is rotating at velocity $\sim v_{\rm k}$, implying that stream elements separated by more than $\sim r_{\rm s}$ collide with ``fresh'' disk patches, unhindered by the stream mass that has previously impacted the disk. The disk's azimuthal motion also imparts lateral momentum to the shocked stream, evacuating the collided stream material away from the instantaneous point of stream-disk intersection. In the limit considered in this section, we shall assume that the optical depth around the collision site is \textit{dominated by recently collided stream material}, rather than an accumulation of shocked stream mass obscuring the emission.

We consider stream segments of mass $\delta m \ll m_{\rm ej,\star}$ spanning an initial length $\delta \ell=\ell (\delta m/m_{\rm ej,\star})$, and collision duration $\delta t_{\rm s} \approx \Delta t_{\rm s}(\delta m/m_{\rm ej,\star})$. The photon diffusion time (in analogy with Eq.~\ref{eq:t_diff_star+d}), assuming the shocked segments expand quasi-spherically is $\delta t_{\rm diff} \approx t_{\rm diff,tot} (\delta m/m_{\rm ej,\star})^{1/2}$.

If we consider a partition of the stream where $\delta t_{\rm s} = \delta t_{\rm diff}$ for each segment, we find
\begin{multline}
    \delta m = m_{\rm ej,\star} \pfrac{t_{\rm diff,tot}}{\Delta t_{\rm s}}^2
    \approx 10^{-5} \, {\rm M_\odot} \\ \eta_{0.03}^2 M_{\bullet,6}^{5/3} \rho_{\rm d,-7}^2 P_{\rm orb,8}^{-3} m_1^{3.7} \tilde{\ell}_{35}^{-2} \,,
\end{multline}
and a corresponding diffusion time
\begin{multline}
    \delta t_{\rm diff} \approx t_{\rm diff,tot}^2/\Delta t_{\rm s} \approx \frac{\kappa m_{\rm ej,\star}}{8\pi c r_{\rm H} \tilde{\ell}} \approx \\
    42 \, {\rm min} \; \rho_{\rm d,-7} \, \eta_{0.03} P_{\rm orb,8}^{-4/3} M_{\bullet,6}^{1/3} m_1^{1.4} \tilde{\ell}_{35}^{-1} \;.
\end{multline}

Assuming quasi-spherical expansion from $r_{\rm s}$ to $v_{\rm k} \delta t_{\rm diff}$, the radiated energy per segment is
\begin{equation}
    \delta E_{\rm rad} \approx \delta m \, v_{\rm k}^2 \pfrac{r_{\rm s}}{v_{\rm k} \delta t_{\rm diff}} \approx \delta m \, v_{\rm k}^2 \pfrac{2\tilde{\ell} \,c/v_{\rm k} }{\kappa m_{\rm ej,\star}/4\pi r_{\rm H}^2}
\end{equation}
and the total radiated energy
\begin{multline}
    E_{\rm rad} \approx \frac{m_{\rm ej,\star}}{\delta m} \delta E_{\rm rad} \approx ( 8\pi \tilde{\ell}  )  \frac{c v_{\rm k}r_{\rm H}^2}{\kappa} \approx \\ 
    5\times 10^{45} \, {\rm erg} \; \tilde{\ell}_{35} m_1^{2/3} M_{\bullet,6}^{1/3} P_{\rm orb,8} \,,
\end{multline}
scales linearly with $P_{\rm orb}$, and independent of $m_{\rm ej,\star}$ and the disk properties. Comparing $E_{\rm rad}$ to the stream's kinetic energy, we find the radiative efficiency
\begin{equation} \label{eq:efficiency_seq_of_collisions}
    \varepsilon_{\rm rad} = \frac{E_{\rm rad}}{m_{\rm ej,\star} v_{\rm k}^2} \approx 0.02 \; \eta_{0.03}^{-1} \rho_{\rm d,-7}^{-1} m_1^{-1.5} M_{\bullet,6}^{-1/3} P_{\rm orb,8}^{7/3}  \,.
\end{equation}
The flare luminosity is then given by
\begin{multline} \label{eq:L_bol_stream_disk_col}
    L_{\rm rad} \approx \frac{E_{\rm rad}}{\Delta t_{\rm s}} \approx L_{\rm Edd} \pfrac{r_{\rm s}}{a_0} \approx \\
    2\times 10^{42} \, {\rm erg \, s^{-1}} \; m_1^{1/3} M_{\bullet,6}^{2/3} \,,
\end{multline}
essentially reproducing a result similar to \cite{Linial_Metzger_23}. Note that $L_{\rm bol}$ is independent of $\tilde{h}$, since the radiating gas is the shocked stream, which is not penetrating the disk.

Again, adhering to the assumption of rapid thermalization, the blackbody temperature is given by
\begin{multline} \label{eq:T_BB_stream_col}
    k_{\rm B} T_{\rm BB} \approx k_{\rm B} \pfrac{L_{\rm rad}}{\sigma_{\rm SB} (v_{\rm k} \,\delta t_{\rm diff})^2}^{1/4}
    \pfrac{c}{v_{\rm k}}^{1/4} \approx \\
    26 \, {\rm eV} \; \pfrac{\tilde{\ell}_{35}}{\eta_{0.03} \rho_{\rm d,-7} M_{\bullet,6}}^{1/2} P_{\rm orb,8}^{11/12}  m_1^{-0.6} \,,
\end{multline}
and as in the wind case, the temperature increases with period and decreasing density, $T_{\rm BB} \propto \rhod^{-1/2} P_{\rm orb}^{11/12}$ (similar scaling as Eq.~\ref{eq:T_BB_wind}).

\section{Radiative Efficiency and the Emission Temperature}
\label{sec:Efficiency}

The radiative efficiency of the stream-disk interaction, $\varepsilon_{\rm rad}$, grows with increasing $P_{\rm orb}$ and decreasing with $\rhod$ (in both emission model flavors, \S \ref{sec:Stream_Wind} and \S \ref{sec:stream_sequence_of_collisions}). The key reason for this trend is that under these conditions, the ejecta mass becomes smaller due to the reduced ram pressure ($m_{\rm ej,\star} \propto \rhod P_{\rm orb}^{-2/3}$, Eq.~\ref{eq:M_ej_Yao}). A smaller ejecta mass implies lower optical depth, allowing radiation generated in the shocked stream to escape more readily, with reduced adiabatic losses.

Naturally, energy conservation imposes $\varepsilon_{\rm rad} \leq 1$. It can be shown that $\varepsilon_{\rm rad} \approx 1$ corresponds to $\tau_{\rm sh} \approx c/v_{\rm k}$, where $\tau_{\rm sh}$ is the optical depth of the mass ahead of the shock (either of the quasi-spherical wind discussed in \S \ref{sec:Stream_Wind}, or the optical depth associated with the relevant stream segment, $\delta m$ discussed in \S \ref{sec:stream_sequence_of_collisions}).

A key assumption underlying our derivations is the the existence of a \textit{radiation mediated shock} (RMS) at the stream-disk interface, effectively converting the stream's kinetic energy into radiation. However, an RMS can only be sustained if photons created at the shock front remain trapped and advected along with the fluid, such that the deceleration of the incoming flow is mediated by the pressure of the resultant radiation field. This condition is met only if the optical depth near the shock front exceeds $\tau_{\rm sh} \approx c/v_{\rm sh} \approx c/v_{\rm k} \approx 10 \, P_{\rm orb,8}^{1/3}M_{\bullet,6}^{-1/3}$ -- which coincides with the criterion for $\varepsilon_{\rm rad} \approx1$. If the optical depth falls below this threshold, newly created photons decouple from the gas and readily escape. In this case, the shock transitions to be \textit{collisionless}, mediated by collective plasma processes rather than radiation pressure (see \citealt{Levinson_Nakar_2020} for a comprehensive review on RMS physics).

Fig.~\ref{fig:M_ej_Period_Diagram} shows the parameter space of stream-disk collisions, represented in terms of the debris mass, $m_{\rm ej,\star}$ versus $P_{\rm orb}$. The choice of $m_{\rm ej,\star}$ as the salient physical parameter, rather than disk properties $\rhod$, eliminates uncertainties and degeneracies concerning the ablation process (e.g., the stripping efficiency, $\eta$, Eq.~\ref{eq:M_ej_Yao}, stellar structure). The stream geometry upon the subsequent disk collision (discussed in \S \ref{sec:Debris_and_stream}) sets the mean density, $\rho_{\rm s}$, or equivalently, the the optical depth in the vicinity of the shock, $\tau_{\rm sh} \approx \kappa \rho_{\rm s} r_{\rm s}$. The blue curve corresponds to $\tau_{\rm sh} \approx c/v_{\rm sh}$ - where the radiative efficiency is $\varepsilon_{\rm rad}\approx1$, and near the boundary between an RMS and a collisionless shock. At yet lower ejecta masses, the maroon curve shows $\tau \approx1$, well-within the collisionless shock regime. The red curve corresponds to where the stream arrival time is comparable to the photon diffusion time of the entire shocked debris mass, $t_{\rm diff} = \Delta t_{\rm s}$ (Eq.~\ref{eq:t_diff_tot_verus_ts}). Our derivations are most relevant in the region between the red and blue curves (i.e., where $t_{\rm diff}(m_{\rm ej,\star}) < \Delta t_{\rm s}$ and $\tau_{\rm sh} > c/v$). In this region, we highlight the parameter space where the blackbody temperature of the shocked-stream emission exceeds $\kb T_{\rm BB} \gtrsim 50 \, {\rm eV}$ (Eq.~\ref{eq:T_BB_stream_col}), and can therefore outshine the Wien-like tail of the inner disk and be detectable as a soft X-ray flare. The dashed gray lines show constant EMRI destruction timescales, $\tau_{\rm life} \approx P_{\rm orb} (m_\star/m_{\rm ej,\star})$.

\begin{figure}
    \centering
    \includegraphics[width=\linewidth]{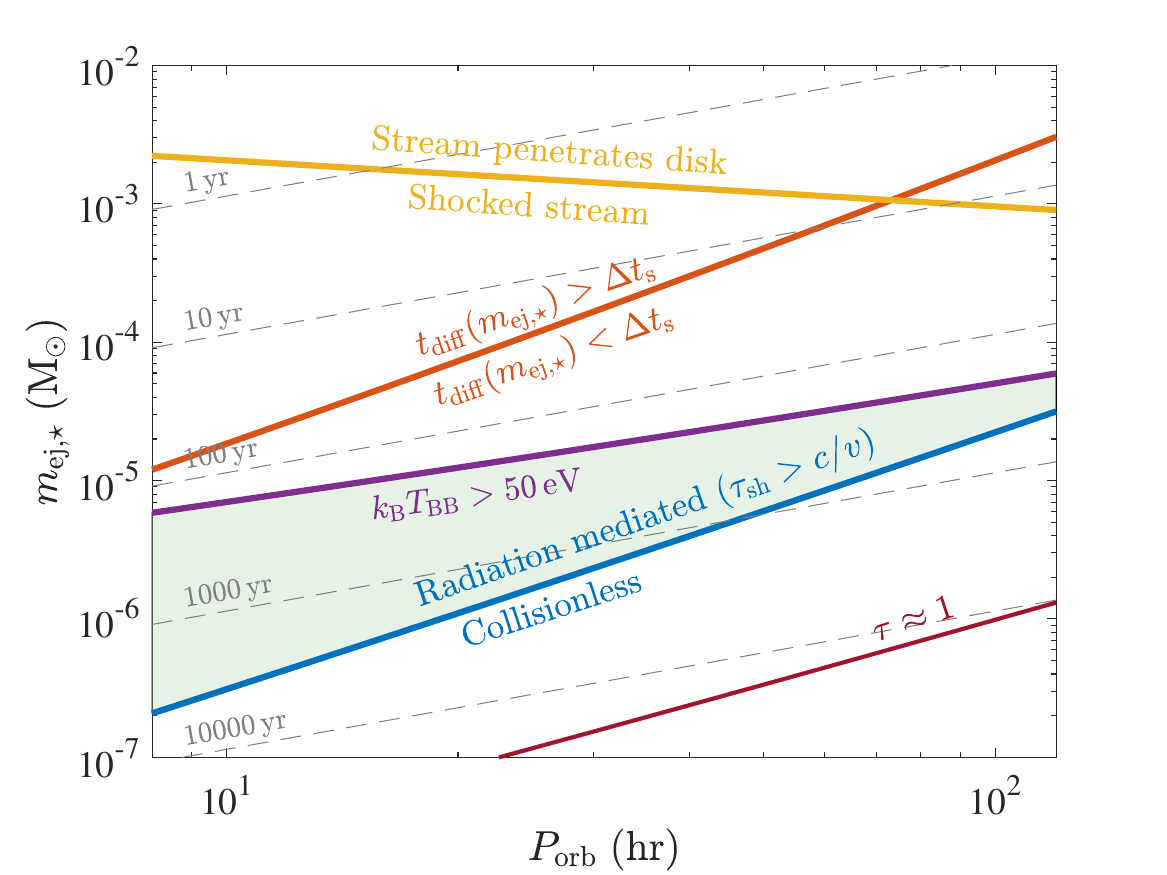}
    \caption{Parameter space of $m_{\rm ej,\star}$ versus $P_{\rm orb}$ for derbis stream+disk interaction (as described in \S \ref{sec:stream_sequence_of_collisions}). At sufficiently low $m_{\rm ej,\star}$, the impacting debris stream is optically thin (maroon line, $\tau\approx1$). When the optical depth ahead of the reverse shock at the stream-disk interface is less than $\tau_{\rm sh} \lesssim c/v$, the shock is collisionless (blue line). At higher masses ($\tau_{\rm sh} \gtrsim c/v$) the shock becomes radiation mediated, with a downstream shock temperature given in Eq.~\ref{eq:T_BB_stream_col}, assuming rapid thermalization. As the mass increases further, the increased optical depth reduces the temperature of the escaping emission, up to the purple line, where $\kb T_{\rm BB} = 50 \, \rm eV$ (similar to the observed temperature of the quiescent QPE emission). At yet higher masses (red line), the interaction becomes ``explosive'', with the flare duration set by the photon diffusion time of the shocked debris rather than the stream arrival time. The yellow line is the (maximal) disk mass participating in the collision assuming a spreading TDE disk, as discussed in Sec.~\ref{sec:OuterDisk}. Above this mass, the stream punches through the disk, and the emission is dominated by shocked disk material. Dashed gray lines correspond to the destruction time of an EMRI of mass $1\, M_\odot$, losing $2m_{\rm ej,\star}$ per orbit. The highlighted region is where X-ray QPEs are likely to be observed.}
    \label{fig:M_ej_Period_Diagram}
\end{figure}

\subsection{Collisionless shocks}

In the absence of an RMS, the gas temperature downstream of the collisionless shock is
\begin{equation}
    \kb T_{\rm sh} \approx (3/16) m_{\rm p} v_{\rm k}^2 \approx 2 \, {\rm MeV} \; (M_{\bullet,6}/P_{\rm orb,8})^{2/3} \,.
\end{equation}
The hot electrons in the shocked downstream cool through bremsstrahlung emission on a timescale $t_{\rm ff} \approx \frac{3}{2} n_{\rm e} \kb T_{\rm sh} / \Lambda(n_{\rm e},T_{\rm sh})$, where $n_{\rm e}$ is the electron density, and $\Lambda(n_{\rm e},T_{\rm sh}) \propto n_{\rm e}^2 T_{\rm sh}^{1/2}$ is the free-free cooling rate. Since $\Lambda$ drops with decreasing density, most of the electron cooling occurs during the time the shocked debris spends at high densities -- $t_{\rm dyn} \approx r_{\rm s}/v_{\rm k}$, before it rarefies and expands quasi-spherically. Comparing the two timescales, we find $t_{\rm ff}/t_{\rm dyn} \propto \rho_{\rm s}^{-1} P_{\rm orb}^{-4/3} \propto \rhod^{-1} P_{\rm orb}^{4/3}$ (where $\rho_{\rm s} \propto \rhod P_{\rm orb}^{-8/3}$, Eq.~\ref{eq:rho_s_relative}). Thus, at sufficiently low $\rhod$ or long $P_{\rm orb}$ (where at face value, $\varepsilon_{\rm rad} \gg 1$), only a small fraction, $t_{\rm dyn}/t_{\rm ff} \ll 1$ of the stream energy is radiated, and the shocked gas evolves almost adiabatically. Since the bremsstrahlung emissivity peaks around $\kb T_{\rm sh}$, most of the radiated energy will at least initially be in $\rm MeV$ energies, with little contribution to soft bands. The resulting spectrum in this regime depends on various factors, including Compton scattering, the Klein-Nishina cross section, beamed emission -- whose treatment is beyond the scope of this paper. Our expectation is that the emission here will generally be hotter than that emitted by an RMS (i.e., $\varepsilon_{\rm rad} \lesssim 1$), with relatively weak flux in soft X-rays. {We note that studies including \cite{Margalit_22}, addressing emission from circumstellar material shocked by a supernova, provide relevant insights to the radiative processes prevailing in collisionless shocks in regimes similar those of stream-disk interaction. We defer more detailed study of the emerging radiation in these regimes to future work.}

\subsection{Radiative Efficiency and temperature}

The luminosity arising from the stream-disk interaction can be expressed as $L_{\rm bol} = \varepsilon_{\rm rad} L_{\rm kin}$, where
\begin{equation}
    L_{\rm kin} \approx \frac{1}{2} \frac{m_{\rm ej,\star}}{\Delta t_{\rm s}} v_{\rm k}^2 \approx \frac{G\MBH m_{\rm ej,\star}}{2a_0 \Delta t_{\rm s}}\,.
\end{equation}
is the stream's kinetic luminosity. The radiative efficiency is ultimately set by the mass content of the stream. Expressing $\varepsilon_{\rm rad}$ in terms of $L_{\rm kin}$, we have
\begin{equation}
    \varepsilon_{\rm rad} \approx
    \begin{cases}
        0.01 \; \pfrac{\Delta \Omega}{\pi}^{2/3} \pfrac{L_{\rm kin}}{L_{\rm Edd}}^{-2/3} m_1^{2/9} M_{\bullet,6}^{-2/9} \; , \; & \S \ref{sec:Stream_Wind}\\
        0.01 \, \pfrac{L_{\rm kin}}{L_{\rm Edd}}^{-1} m_1^{1/3} M_{\bullet,6}^{-1/3} \; , \; & \S \ref{sec:stream_sequence_of_collisions}
    \end{cases}
\end{equation}
where the two cases correspond to the ``wind-like'' emission (Eq.~\ref{eq:efficiency_wind_params}) and the sequence of collisional ``explosions'' (Eq.~\ref{eq:efficiency_seq_of_collisions}), respectively. The bolometric luminosities can be derived in terms of $\varepsilon_{\rm rad}$
\begin{multline}
    L_{\rm bol} \approx L_{\rm kin}(\varepsilon_{\rm rad}) \varepsilon_{\rm rad} \approx \\
    \begin{cases}
        1.5\times 10^{42} \, {\rm erg \, s^{-1}} \,M_{\bullet,6}^{2/3} m_1^{1/3} \varepsilon_{\rm rad}^{-1/2} (\Delta\Omega/\pi) \; , \; & \S \ref{sec:Stream_Wind} \\
        2\times 10^{42} \, {\rm erg \, s^{-1}} \, M_{\bullet,6}^{2/3} m_1^{1/3} \; , \; & \S \ref{sec:stream_sequence_of_collisions}
    \end{cases}
\end{multline}
where we used Eqs.~\ref{eq:L_bol_stream_disk_wind} and \ref{eq:L_bol_stream_disk_col}\footnote{At a fixed orbital period, high radiative efficiency corresponds to less optical depth, lower $m_{\rm ej,\star}$, and thus to lower $L_{\rm kin}$. This leads to the somewhat counterintuitive result $L_{\rm rad} \propto \varepsilon_{\rm rad}^{-1/2}$, where the absolute radiated luminosity increases with decreasing $\varepsilon_{\rm rad}$.}. Thus, for a fixed $\varepsilon_{\rm rad}$, we find $L_{\rm bol}$ is \textit{independent} of the disk properties and orbital period ($\rhod$ and $P_{\rm orb}$), despite the non-trivial dependence on these parameters in the form given earlier in Eq.~\ref{eq:L_bol_stream_disk_wind}.

Finally, assuming rapid thermalization is achieved, we express the emission blackbody temperature as a function of $\varepsilon_{\rm rad}$
\begin{multline} \label{eq:T_BB_stream_eff}
    \kb T_{\rm BB} = \kb \pfrac{\rho_{\rm s} v_{\rm k}^2}{a_{\rm rad}}^{1/4} \times 
    \begin{cases}
        \varepsilon_{\rm rad} \; , \; \S \ref{sec:Stream_Wind} \\
        \varepsilon_{\rm rad}^{3/4} \; , \; \S \ref{sec:stream_sequence_of_collisions}
    \end{cases} \\
    \approx\begin{cases}
        200 \, {\rm eV} \; \varepsilon_{\rm rad}^{5/8} P_{\rm orb,8}^{-1/4} (\Delta \Omega/\pi)^{2/3} M_{\bullet,6}^{1/12} m_1^{-1/12} \; , \; &\S \ref{sec:Stream_Wind} \\
        200 \, {\rm eV} \; \varepsilon_{\rm rad}^{1/2} P_{\rm orb,8}^{-1/4} M_{\bullet,6}^{1/12} m_1^{-1/12} \; , \; &\S \ref{sec:stream_sequence_of_collisions} 
    \end{cases}
\end{multline}
following Eqs.~\ref{eq:T_BB_wind} and \ref{eq:T_BB_stream_col},  and where we used $L_{\rm kin} = \pi r_{\rm s}^2 \rho_{\rm s} v_{\rm s}^3/2$. Thus, for a fixed $\varepsilon_{\rm rad} \lesssim 1$, the emitted temperature scales relatively weakly with orbital period as $T_{\rm BB} \propto P_{\rm orb}^{-1/4}$. 

\subsection{High radiative efficiency}

Radiative efficiency of $\varepsilon_{\rm rad}\lesssim 1$ corresponds to an optical depth $\tau \gtrsim c/v_{\rm sh}$, allowing for a radiation mediated shock to develop at the stream-disk collision interface. In the marginal case of $\varepsilon_{\rm rad} \approx 1$, the total emitted energy is comparable to the stream's kinetic energy, $E_{\rm rad} \approx E_{\rm kin} =m_{\rm ej,\star} v_{\rm k}^2 / 2$. Since the shock front is only mildly obstructed, the emitted temperature closely traces the downstream temperature of the RMS. Assuming rapid thermalization \citep{Vurm_25}, the blackbody temperature predicted by Eq.~\ref{eq:T_BB_stream_eff} is consistent with observed QPE flare temperatures when $\varepsilon_{\rm rad} \sim 1$.

At yet lower optical depths, $\tau < c/v_{\rm sh}$, an RMS cannot be sustained and the stream is decelerated and deflected by a collisionless shock instead. The shocked stream is heated to gas temperatures of $\kb T_{\rm sh} \approx \rm (0.1-1) \,MeV$. When $\tau \ll c/v_{\rm sh}$, the shock is radiatively inefficient -- the short time spent at the highest densities in the immediate downstream (where photon production is dominant) precludes efficient cooling through free-free emission and most of the energy remains in the form of hot, slowly cooling gas, similarly to the adiabatic regime discussed in \cite{Margalit_22}. We will address the observational signatures of dilute streams impacting the disk in future work.

Consider an EMRI at a fixed $P_{\rm orb}$. At very low disk densities, $\rhod \ll 10^{-9} \, {\rm g \, cm^{-3}} \, P_{\rm orb,8}^{7/3}$, the resulting shock is collisionless producing a weak hard X-ray/gamma-ray signal, with a low-energy tail that likely has a fairly weak soft X-ray flux. In contrast, at densities $\rhod \gg 10^{-9} \, {\rm g \, cm^{-3}} \, P_{\rm orb,8}^{7/3}$ the shock is radiation mediated, with a relatively soft emission temperature scaling as $\kb T_{\rm BB} \propto \rhod^{-1/2}$ (as in Eq.~\ref{eq:T_BB_stream_col}) and peaking well below the observed X-ray band. A density scale therefore exists satisfying $\varepsilon_{\rm rad} \sim 1$ (Eq.~\ref{eq:efficiency_seq_of_collisions})
\begin{equation} \label{eq:rho_eff_1}
    \rho_{\rm d}^{\varepsilon_{\rm rad}\approx 1} \approx 2\times10^{-9} \, {\rm g \, cm^{-3}} \; \eta_{0.03}^{-1} m_1^{-1.5} M_{\bullet,6}^{-1/3} P_{\rm orb,8}^{7/3} \,,
\end{equation}
at which point the resulting emission peaks in soft X-rays. This suggests that X-ray QPEs may appear if over the course of the disk evolution, densities comparable to $\sim \rho_{\rm d}^{\varepsilon_{\rm rad}\approx1}$ are realized. We consider feasible disk densities and the resulting emission properties in the next section.

\section{Accretion disk parameters} \label{sec:DiskModels}

Thus far, we have been agnostic regarding the nature of the underlying disk, adopting arbitrary fiducial values of $\rhod$ and $\tilde{h}$ in our derivations. Here we consider more concrete (albeit simplified) disk models, and specialize previous results. Specifically, we are interested in two limits of the EMRI+disk interaction:
\begin{enumerate}
    \item The EMRI resides well within the disk's outer edge, $a_0 \ll R_{\rm d}$. The EMRI then encounters a quasi-steady accretion flow, with radially uniform $\dot{M}_{\rm d}$, slowly evolving over the disk's exterior viscous timescale. This limit is appropriate for describing either an extended AGN-like disk or the interior of a slowly spreading TDE disk, extending past the EMRI's orbit.
    \item The EMRI is impacting disk material near its outer-edge, i.e., $a_0 \lesssim R_{\rm d}$. This could occur if the EMRI is positioned well-outside the TDE circularization radius, such that EMRI-disk collisions commence after the disk has undergone sufficient spreading to intercept the EMRI orbit. In this limit, the disk surface density $\Sigmad$ rises to a peak (over the disk spreading timescale), followed by a gradual decay.
\end{enumerate}

Joint X-ray/UV observations of the quiescent emission in QPEs have been used to constrain the size and evolution of their underlying accretion disks, motivating both of the above regimes \citep{Nicholl_24,Wevers_2025_ero2, Guolo_2025_GSN069_time_dependent,Chakraborty_25_22upj}. In the long-period QPEs ($\left<P_{\rm QPE} \right> \gtrsim 1\,\rm d$) AT 2019qiz and AT 2022upj, QPEs were detected 3-4 years following the detection of an optical TDE in the same system. In these sources, the position of an orbiter of period $P_{\rm orb} = \left< P_{\rm QPE} \right>$ is consistent with the outer edge of fitted spreading-disk model, i.e., $a_0 \approx R_{\rm d}$. On the other hand, in short period sources ($\left<P_{\rm QPE} \right> \approx \mathcal{O}(\rm hrs)$), GSN 069 and eRO-QPE2, the orbiter appears to be well within the inner region of the disk, $a_0 \ll R_{\rm d}$.

\subsection{Steadily accreting disk} \label{sec:SteadyDisk}

For a given disk accretion rate $\dot{M}_{\rm d}$, the surface density encountered by the EMRI is \citep[e.g.,][]{Frank+02}
\begin{multline}
    \Sigmad = \frac{\dot{M}_{\rm d}}{3\pi \nu} =
    4\times 10^5 \, {\rm g \, cm^{-2}} \frac{\lambda_{\rm Edd}}{\alpha_{-1} \tilde{h}_{-2}^2} \pfrac{M_{\bullet,6}}{P_{\rm orb,8}}^{1/3}
\end{multline}
where $\lambda_{\rm Edd} = \dot{M}_{\rm d}/\dotMEdd$ and $\nu$ is the disk viscosity, parameterized as $\nu= \alpha\sqrt{G\MBH r}(\Hd/r)^2$, where $\alpha \approx 0.1 \alpha_{-1}$ is the Shakura-Sunyaev alpha parameter \citep{Shakura1976}. The disk's midplane density is given by
\begin{multline} \label{eq:rho_d_steady}
    \rhod \approx \frac{\Sigmad}{\Hd}
    \approx 2.6\times 10^{-6} \, {\rm g \, cm^{-3}}
    \frac{\lambda_{\rm Edd}}{\alpha_{-1} P_{\rm orb,8}} \tilde{h}_{-2}^{-3} \,,
\end{multline}
strongly dependent on $\tilde{h}$, which we leave as a free parameter for the time being.

\subsection{Outer edge of a spreading disk} \label{sec:OuterDisk}
Consider a disk formed by a second star of mass and radius $m_\star^{\rm tde}$, $R_\star^{\rm tde}$, undergoing a TDE as it approaches the SMBH on a parabolic trajectory with pericenter distance $r_{\rm p}^{\rm tde} \approx R_\star^{\rm tde} (\MBH/m_{\star}^{\rm tde})^{1/3}$. If angular momentum is conserved as the bound debris circularizes, the disk forms with an initial radius $R_{\rm d,0} \approx 2 r_{\rm p}^{\rm tde}$ (although see caveats and challenges to this picture discussed in e.g., \citealt{Lu_Bonnerot_2020}). The orbital period at the circularization radius is given by the same expression\footnote{Both the circularization radius and the Roche limit are approximately twice the tidal radius, but for different reasons. The TDE circularization radius is given by angular momentum conservation, with the specific angular momentum of the TDE progenitor being $\sqrt{G\MBH (1+e) r_{\rm p}^{\rm tde}} \approx \sqrt{2G\MBH \rtidal^{\rm tde}}$. The minimal semi-major axis of the stellar EMRI, $\approx 2\rtidal$, is a consequence of the Roche lobe radius at the limit $m_\star/\MBH\ll1$, $\approx 0.46 \, a_0 (m_\star/\MBH)^{1/3}$ \citep[e.g.,][]{Eggleton_83}.} as Eq.~\ref{eq:P_orb_min}, specialized for $m_\star = m_\star^{\rm tde}$. 

If the EMRI is orbiting far outside $2r_{\rm p}^{\rm tde}$, the interaction with the TDE disk commences after it has spread considerably, when $R_{\rm d} \approx a_0$. The surface density encountered by the EMRI evolves with time according to $\Sigmad(t;a_0)$ -- given by the Green's function of the one-dimensional disk diffusion equation \citep[e.g.,][]{Pringle_72,Cannizzo_1990,Mummery_2024}. The disk surface density at a given radius rises to peak around $a_0 \approx R_{\rm d}$, followed by a gradual decay as the outer edge continues to spread further to $R_{\rm d} \gg a_0$, with $M_{\rm d}$ decaying with time.

With its net angular momentum conserved, the remaining disk mass scales with $R_{\rm d}$ as
\begin{multline} \label{eq:M_disk_spreading}
    M_{\rm d} = M_{\rm d,0} (R_{\rm d}/R_{\rm d,0})^{-1/2} = M_{\rm d,0} P_{\rm orb,8}^{-1/3} (m_{\star,1}^{\rm tde})^{0.23} = \\
    1 \, {\rm M_\odot} \; f_{\rm d} P_{\rm orb,8}^{-1/3} (m_{\star,1}^{\rm tde})^{1.23}  \,,
\end{multline}
where $m_{\star,1}^{\rm tde} = m_\star^{\rm tde}/\rm M_\odot$, and we assumed that the initial disk mass is a fraction $f_{\rm d}\lesssim 0.5$ of the disrupted star's mass, $M_{\rm d,0} = f_{\rm d} m_{\star}^{\rm tde}$.

The maximal surface density encountered by the EMRI is therefore, up to some order unity prefactor
\begin{multline} \label{eq:Sigma_d_spreading}
    \max\{\Sigmad(a_0)\} \approx \frac{M_{\rm d}}{\pi R_{\rm d}^2} = \frac{M_{\rm d,0}}{4 \pi \rtidal^2} \pfrac{M_{\rm d}}{M_{\rm d,0}} \pfrac{R_{\rm d}}{2\rtidal}^{-2}
    \approx\\ 3.3\times 10^{6} \, {\rm g \, cm^{-2}} \; f_{\rm d} P_{\rm orb,8}^{-5/3} M_{\bullet,6}^{-2/3} (m_{\star,1}^{\rm tde})^{1.23} \,,
\end{multline}
and the corresponding mass density
\begin{multline} \label{eq:rho_d_spreading}
    \rhod=\frac{\max\{\Sigmad(a_0)\}}{\Hd} \approx \pfrac{M_{\rm d}}{\MBH} \frac{4\pi}{G P_{\rm orb}^2} \pfrac{\Hd}{R_{\rm d}}^{-1} = \\
    2.3\times 10^{-5} \, {\rm g \, cm^{-3}} \, f_{\rm d} P_{\rm orb,8}^{-7/3} (m_{\star,1}^{\rm tde})^{1.23} ( \tilde{h}_{-2} M_{\bullet,6} )^{-1} \,.
\end{multline}

At a fixed position $a_0$, the surface density can be expressed as $\Sigmad = \eta_\Sigma \max\{ \Sigmad(a_0) \}$ where $\eta_\Sigma  < 1$ is a time-dependent prefactor, reflecting the increase and decay of $\Sigmad$ before and after peak. As a concrete example, in the case of a radiatively cooling, gas-pressure dominated disk with constant opacity (as studied in e.g., \citealt{Cannizzo_1990}), $\eta_\Sigma \approx 7 (t/t_0)^{-57/80} (1-(t/t_0)^{-21/40})^{3/2}$ where $t_0$ is the time when $R_{\rm d} = a_0$, and peak $\Sigmad$ occurs at around $t\approx4 t_0$, with $\eta(t\approx 4\,t_0) \approx 1$.

As a caveat, we note that the disk formation process and its evolution in early times is subject to various theoretical uncertainties (see, e.g., \citealt{Lu_25_tde_disks} for review). For example, during the first ${\sim}\rm yr$ following the disruption, fallback rates of bound debris exceed the Eddington limit. At this stage the disk is highly inflated ($\tilde{h}\sim0.3$) and may undergo rapid viscous spreading while losing substantial mass to outflows, suggesting $f_{\rm d} \ll 0.5$ \citep{Metzger&Stone16,Jiang_2019}. Given these and other uncertainties, we conservatively limit our discussion of the $a_0 \approx R_{\rm d}$ limit to disks that have expanded well-beyond $2\rtidal$. For $m_\star^{\rm tde} = 1 \, \rm M_\odot$, we take $P_{\rm orb} = 24 \, {\rm hr} \, P_{\rm orb,24}$ as our fiducial period. Fig.~\ref{fig:TimescaleComparison} shows in dashed colored lines the photon diffusion timescale of the various emission components, assuming a spreading TDE disk, as described here.

\subsection{Constraints on disk mass and angular momentum}

We briefly address the compatibility of the two regimes considered here. The total mass content of an accretion disk is dominated by its outer radii, i.e., $\Sigmad a_0^2$ is an increasing function of $a_0$ for reasonable assumptions about the disk physics. Considering a steadily accreting disk of Eddington ratio $\lambda_{\rm Edd}$, the mass enclosed within the EMRI orbit is at most
\begin{equation} \label{eq:M_steady_enc}
    M_{\rm d}(\lesssim a_0;\lambda_{\rm Edd}) \lesssim \pi a_0^2 \Sigmad \approx 0.13 \, {\rm M_\odot} \frac{\lambda_{\rm Edd}}{\alpha_{-1} \tilde{h}_{-2}^2} M_{\bullet,6}^{2/3} P_{\rm orb,8}^{1/3} \,.
\end{equation}
If it indeed describes the inner region of a slowly evolving TDE disk, the enclosed mass $M_{\rm d}(\lesssim a_0 ;\lambda_{\rm Edd})$ cannot exceed the remaining TDE disk mass (Eq.~\ref{eq:M_disk_spreading}), with its outer radius outside the EMRI orbit ($R_{\rm d}> a_0$). Equating the two expressions, the critical period where the mass of a steady disk matches the remaining TDE disk mass is $P_{\rm orb}(M_{\rm d}(\lambda_{\rm Edd})=M_{\rm d}^{\rm tde})\approx 7.5 \, {\rm d} \, (f_{\rm d} \alpha_{-1}/\lambda_{\rm Edd})^{3/2} (m_{\star,1}^{\rm tde})^{1.8} \tilde{h}_{-2}^3 / M_{\bullet,6} \,.$

As a concrete example, we consider a simple model, of a radiatively cooling alpha-disk with $\alpha=0.1$ around an SMBH of mass $M_\bullet=10^6 \, \rm M_\odot$, with constant opacity $\kappa=\kappa_{\rm es}$, supported by gas and radiation pressure \citep{Shakura1976}. We solve for the aspect ratio $\tilde{h}$ as a function of $P_{\rm orb}$ for a fixed Eddington ratio $\lambda_{\rm Edd}$, and compute the enclosed mass $M_{\rm d}(<P_{\rm orb})=\int^{a_0} 2\pi r\Sigmad(r)dr$. The results are plotted in Fig.~\ref{fig:EnclosedMassSteadyDisk}, showing the enclosed disk mass for different values of $\lambda_{\rm Edd}$ (colored solid curves). The gradual transition between two power laws regimes (seen e.g., for $\lambda_{\rm Edd} = 10^{-2}$), is due to the dominance of gas vs. radiation pressure. The black curves show the remaining TDE disk mass (Eq.~\ref{eq:M_disk_spreading}) as its outer edge has expanded to period $P_{\rm orb}$, assuming $f_{\rm d}=0.5$ and for different values of $m_\star^{\rm tde}$. The plot demonstrates that for $P_{\rm orb} \lesssim \mathcal{O}(\rm 1 \, d)$, the enclosed mass of our simple Shakura-Sunyaev disk is always smaller than the remaining mass of a TDE disk which has expanded to the same period (for $m_\star^{\rm tde} \gtrsim1 \, \rm M_\odot$). At longer periods, as seen in ``Ansky'', $P_{\rm orb} > 5 \, \rm d$, a steadily accreting disk with $\dot{M}_{\rm d}$ in the range of observed QPE quiescent emission, is only compatible with higher $m_\star^{\rm tde}$, or alternatively, with a yet more extended AGN disk, not subject to the TDE mass/angular momentum constraint \citep[e.g.,][]{Mummery_25}.

It is worth noting that the alpha-disk model considered here likely overestimates the surface and mass densities of realistic steadily accreting disks. Magnetic pressure support and higher disk opacities tend to inflate the vertical scale height relative to the simple estimates used here \citep[e.g.,][]{Jiang_16_iron,Zhang_2025}. Consequentially, for the same $\lambda_{\rm Edd}$, enclosed disk masses are likely well-below what is plotted in Fig.~\ref{fig:EnclosedMassSteadyDisk} (recall that $M_{\rm d}(\lesssim a_0) \propto \tilde{h}^{-2}$, Eq.~\ref{eq:M_steady_enc}). The limits imposed by the mass/angular momentum budget of TDE disks are therefore only met at much longer periods (i.e., the intersection of the colored and black curves in Fig.~\ref{fig:EnclosedMassSteadyDisk} occurs at longer periods).

\begin{figure}
    \centering
    \includegraphics[width=\linewidth]{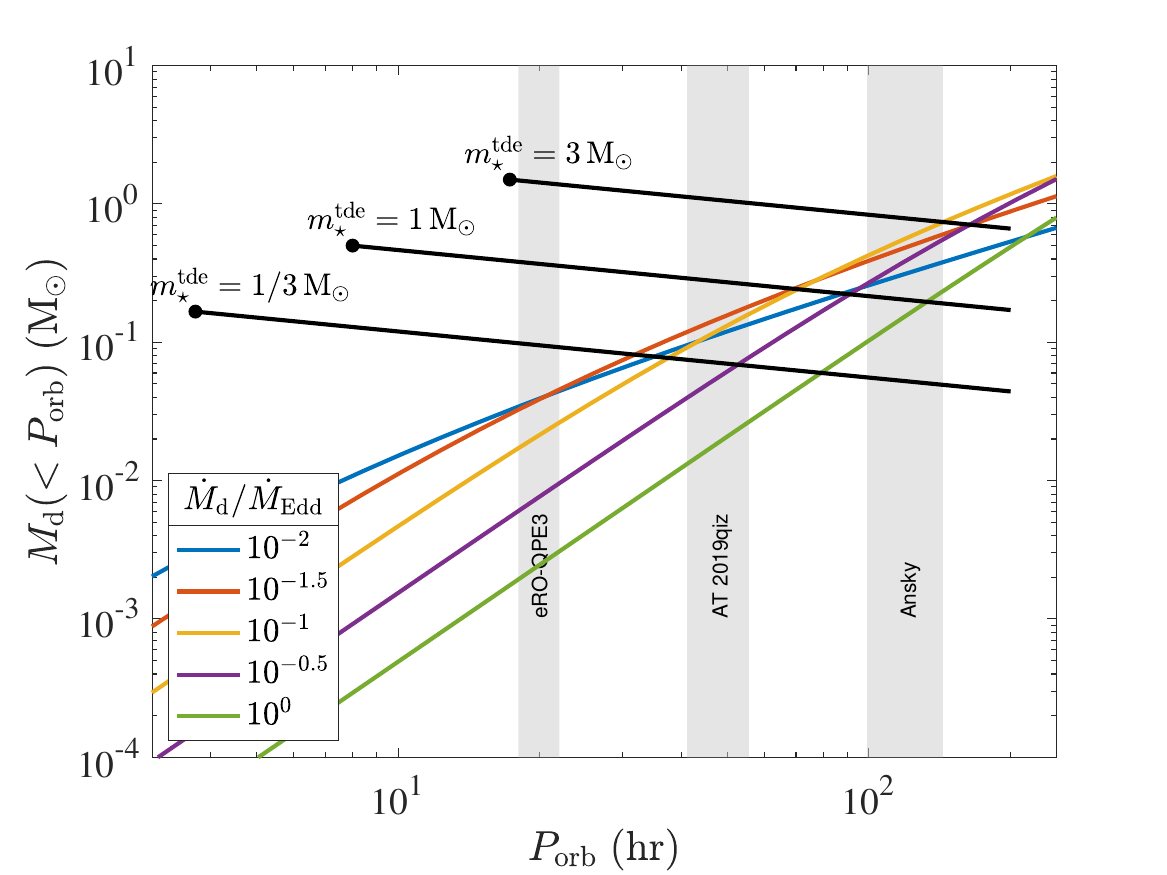}
    \caption{The disk mass of a steadily accreting alpha-disk \citep{Shakura1976} around $M_\bullet=10^6 \, \rm M_\odot$, with $\alpha=0.1$ and constant opacity $\kappa=\kappa_{\rm es}$. Solid colored curves correspond to different accretion rates relative to the Eddington rate. The black solid curves show the TDE disk mass for different progenitor masses, $m_\star^{\rm tde}$, as it viscously spreads with the orbital period at its outer edge growing to $P_{\rm orb} \gtrsim P_{\rm orb}(2\rtidal)$, assuming $f_{\rm d}=0.5$. The highlighted strips show the flare recurrence times for a few QPE sources. The intersection of the black and colored lines show the maximum $\dot{M}$ supplied by a TDE of the star of the given mass.}
    \label{fig:EnclosedMassSteadyDisk}
\end{figure}

\begin{deluxetable*}{lccc}
\tablecaption{Summary of timescales, luminosities, and blackbody temperatures for different interaction models, assuming the EMRI collides with the inner region of a steadily accreting disk.\label{tab:inner_disk}}
\tablehead{
    \colhead{} & 
    \colhead{$\Delta t$ (hr)} & 
    \colhead{$L_{\rm bol}$ (erg s$^{-1}$)} & 
    \colhead{$k_{\rm B} T_{\rm BB}$ (eV)}
}
\startdata
Star+disk & 
$0.5 \; R_1 \left(\frac{\lambda_{\rm Edd}}{\alpha_{-1}}\right)^{1/2} \tilde{h}_{-2}^{-1}$ & 
$10^{42} \; R_1^{2/3} \tilde{h}_{-2}^{1/3} M_{\bullet,6}^{7/9} P_{\rm orb,8}^{-4/9} $ & 
$9 \; \pfrac{\alpha_{-1}}{\lambda_{\rm Edd}}^{1/4} \tilde{h}_{-2}^{7/12} M_{\bullet,6}^{-1/18}P_{\rm orb,8}^{5/36} R_1^{-1/3} $ \\
Hill sphere+disk & 
$1 \; P_{\rm orb,8} \left(\frac{\lambda_{\rm Edd}}{\alpha_{-1}}\right)^{1/2} \tilde{h}_{-2}^{-1} m_1^{1/3} M_{\bullet,6}^{1/6}$ & 
$1.5\times10^{42} \; \tilde{h}_{-2}^{1/3} M_{\bullet,6}^{7/9} m_1^{2/9}$ & 
$7 \; \pfrac{\alpha_{-1}}{\lambda_{\rm Edd}}^{1/4} \tilde{h}_{-2}^{7/12} M_{\bullet,6}^{-5/36} P_{\rm orb,8}^{-1/4} m_1^{-1/9}$ \\
Stream+disk (wind) & 
$0.9 \; P_{\rm orb,8} (m_1/M_{\bullet,6})^{1/3}$ & 
\makecell[l]{$3\times10^{43} \;
m_1^{0.96} P_{\rm orb,8}^{-10/9}$ \\ $M_{\bullet,6}^{8/9} (\lambda_{\rm Edd}/\alpha_{-1})^{1/3} \tilde{h}_{-2}^{-1}$} & 
{$2 \pfrac{\alpha_{-1}}{\lambda_{\rm Edd}}^{5/12} \tilde{h}_{-2}^{5/4} M_{\bullet,6}^{-2/3}P_{\rm orb,8}^{41/36} m_1^{-0.7}$} \\
Stream+disk (col.) & 
$0.9 \; P_{\rm orb,8} (m_1/M_{\bullet,6})^{1/3}$ & 
$2\times 10^{42} \; m_1^{1/3} M_{\bullet,6}^{2/3}$ & 
$5 \pfrac{\alpha_{-1}}{\lambda_{\rm Edd}}^{1/2} \tilde{h}_{-2}^{3/2} M_{\bullet,6}^{-1/2} P_{\rm orb,8}^{17/12} m_1^{-0.6}$ \\
\enddata
\end{deluxetable*}

\begin{deluxetable*}{lccc}
\tablecaption{Same as Table~\ref{tab:inner_disk}, but for the case when the EMRI collides with the outer edge of a viscously spreading TDE disk, where $a_0\approx R_{\rm d}$.\label{tab:outer_disk}}
\tablehead{
    \colhead{} & 
    \colhead{$\Delta t$ (hr)} & 
    \colhead{$L_{\rm bol}$ (erg s$^{-1}$)} & 
    \colhead{$k_{\rm B} T_{\rm BB}$ (eV)}
}
\startdata
Star+disk & 
$0.7 \;  P_{\rm orb,24}^{-2/3} R_1 (f_{\rm d} \eta_\Sigma/M_{\bullet,6})^{1/2} (m_{\star,1}^{\rm tde})^{0.6}$ & 
$6\times 10^{41} \; P_{\rm orb,24}^{-4/9}  R_1^{2/3} \tilde{h}_{-2}^{1/3} M_{\bullet,6}^{7/9}$ & 
$9 \; (\eta_\Sigma f_{\rm d})^{-1/4} R_1^{-1/3} (m_{\star,1}^{\rm tde})^{-0.15} \tilde{h}_{-2}^{1/3} M_{\bullet,6}^{7/36} P_{\rm orb,24}^{1/12}$ \\
Hill sphere+disk & 
$3 \; m_1^{1/3} (f_{\rm d} \eta_\Sigma/M_{\bullet,6})^{1/2} (m_{\star,1}^{\rm tde})^{0.6}$ & 
$1.5\times10^{42} \; \tilde{h}_{-2}^{1/3} M_{\bullet,6}^{7/9} m_1^{2/9}$ & 
$5 \; (\eta_\Sigma f_{\rm d})^{-1/4} m_1^{-1/9} (m_{\star,1}^{\rm tde})^{-0.15} \tilde{h}_{-2}^{1/3} M_{\bullet,6}^{7/36} P_{\rm orb,24}^{1/4}$ \\
Stream+disk (wind) & 
$2.7 \; P_{\rm orb,24} \, (m_1/M_{\bullet,6})^{1/3}$ &
\makecell[l]{$10^{43} \;
(f_{\rm d}\eta_\Sigma/\tilde{h}_{-2})^{1/3} m_1^{0.96}$ \\ $P_{\rm orb,24}^{-14/9}M_{\bullet,6}^{5/9} (m_{\star,1}^{\rm tde})^{0.4}$} & 
$5 \; (f_{\rm d} \eta_\Sigma/\tilde{h}_{-2})^{-5/12}P_{\rm orb,24}^{61/36} m_1^{-0.7} (m_{\star,1}^{\rm tde})^{-0.5} M_{\bullet,6}^{-1/4}$ \\
Stream+disk (col.) & 
 $2.7 \; P_{\rm orb,24} \, (m_1/M_{\bullet,6})^{1/3}$ &
$2\times 10^{42} \; m_1^{1/3} M_{\bullet,6}^{2/3}$ &
$17 \; (f_{\rm d} \eta_\Sigma/\tilde{h}_{-2})^{-1/2}P_{\rm orb,24}^{25/12} m_1^{-0.6} (m_{\star,1}^{\rm tde})^{-0.6} $ \\
\enddata
\end{deluxetable*}

\subsection{Emission properties}

The duration, luminosity, and blackbody temperature (assuming efficient thermalization) of the different emission components are summarized in the following tables. Table~\ref{tab:inner_disk} gives expressions and scalings assuming the EMRI interacts with the inner region of a steadily accreting disk (\S \ref{sec:SteadyDisk}). Table~\ref{tab:outer_disk} shows results appropriate for an EMRI crossing near the outer edge of a viscously spreading TDE disk, $a_0 \approx R_{\rm d}$ (\S \ref{sec:OuterDisk}). 

The range of bolometric luminosities subtended by the various emission models presented here is in rough agreement with observed QPE luminosities, $L_{\rm bol} \approx 10^{42-43} \, \rm erg \, s^{-1}$. Yet, as a function of $P_{\rm orb}$, only ``Hill sphere+disk'' and the ``collisional stream+disk'' models predict $L_{\rm bol} \approx \rm const$, for both $a_0 \ll R_{\rm d}$ and $a_0 \approx R_{\rm d}$. In other models, $L_{\rm bol}$ decreases with increasing $P_{\rm orb}$, in tension with the observational trends (\citealt{Arcodia_2024_3&4}, and see also remarks in \S \ref{sec:Intro}). Fixing other parameters, the flare duration increases with $P_{\rm orb}$ in all stream+disk models (for which duration is dominated by $\Delta t_{\rm s}$, the stream arrival time, Eq.~\ref{eq:t_stream_duration}). This is also the case for the diffusion time of disk material shocked by dense ejecta within the Hill sphere, for a steadily accreting inner accretion disk (of fixed $\lambda_{\rm Edd}$ and $\tilde{h}$). Thus, the observed trends, $\Delta t \propto P_{\rm orb}$, $E_{\rm rad} \approx L_{\rm bol} \Delta t \propto P_{\rm orb}$ are satisfied by 3 of the 8 models: Stream+disk collisional model (for both inner and outer disks), and in the Hill sphere+disk model, for an inner, steadily accreting disk.

The blackbody temperatures predicted in all models are generally too low to explain the observed soft X-ray QPE flares for the assumed fiducial values. Photon starvation effects discussed in \cite{Linial_Metzger_23,Vurm_25} may cause the effective emission temperature to exceed $\kb T_{\rm BB}$. This occurs due to the limited rate of photon production in the shocked gas, unable to reach the photon number density dictated by blackbody equilibrium. As the same radiation energy density is shared between fewer photons, the emission is effectively harder than $\kb T_{\rm BB}$. These effects can indeed generate soft X-ray flares, with a partially Comptonized bremsstrahlung spectrum \citep{Nakar_Sari_10}, whose high-energy exponential cutoff is consistent with the observed Wien-like tail, of a quasi-thermal spectrum \citep{Miniutti_23a,Vurm_25}.

Nonetheless, even if blackbody equilibrium is achieved promptly, stream+disk models give rise to soft X-rays within the allowed parameter space. For example, for an interior of a steady disk ($a_0 \ll R_{\rm d}$), stream-disk collisions produce blackbody temperatures of
\begin{multline}
    \kb T_{\rm BB} \approx 200 \, {\rm eV} P_{\rm orb,24}^{17/12} \pfrac{\lambda_{\rm Edd}}{0.1}^{-1/2} \pfrac{\tilde{h}}{0.02}^{3/2} \\
    \pfrac{\alpha_{-1}}{M_{\bullet,6}}^{1/2} m_1^{-0.6} \,.
\end{multline}
without invoking hardening due to photon starvation.
Assuming the collisional stream+disk picture (\S \ref{sec:stream_sequence_of_collisions}), and when the EMRI interacts with the \textit{outer edge of a spreading disk}, soft X-ray flares with $\kb T_{\rm BB} \approx 100 \, \rm eV$ are produced, for example, if the disk is relatively inflated and/or dilute disks. For example, the following combination of parameters
\begin{multline}
    \kb T_{\rm BB} \approx 107 \, {\rm eV} \; \pfrac{f_{\rm d}}{0.5}^{-1/2} \pfrac{\eta_\Sigma}{0.5}^{-1/2} \tilde{h}_{-1}^{1/2} \\
    P_{\rm orb,24}^{25/12} \, m_1^{-0.6} (m_{\star,1}^{\rm tde})^{-0.6} \,,
\end{multline}
yields temperatures that are commensurate with observed QPE flare temperatures.

At any $P_{\rm orb} \gtrsim 1 \, \rm d$ and for fixed $m_\star$ and $m_\star^{\rm tde}$, there is therefore a value of $\eta_\Sigma < 1$ for which a fixed blackbody temperature of $\kb T_{\rm BB}=100 \, {\rm eV}$ is attained. Specifically, in the collisional stream+disk picture, this occurs when
\begin{equation} \label{eq:eta_Sigma_X}
    \eta_\Sigma^{\rm X}(100 \, {\rm eV})\approx 0.06 \; \frac{\tilde{h}_{-2} P_{\rm orb,24}^{25/6}}{(m_1 m_{\star,1}^{\rm tde})^{1.2}} \pfrac{f_{\rm d}}{0.5}^{-1} \,.
\end{equation}
with almost identical scalings for the wind picture. This suggests that as the disk spreads to encounter the EMRI, and $\eta_\Sigma$ increases from 0 to 1 over the disk spreading timescale, $t_{\rm disk}$, soft X-rays are generated on an interval where $\eta_\Sigma\approx\eta_\Sigma^{\rm X}$. Thus, the time spent around a fixed $\kb T_{\rm BB}$ is (up to an order unity factor), approximately $\eta_\Sigma^{\rm X} t_{\rm disk}$ during the gradual increase in disk density encountered by the EMRI. We discuss further implications for QPE detectability in \S \ref{sec:QPE_detectability}.

\section{Discussion and Summary} \label{sec:Discussion}

When a star repeatedly collides with a geometrically thin accretion disk around an SMBH, a small fraction of the stellar mass is stripped through ablation with every disk crossing. We studied the dynamics and observational consequences of this process assuming a main-sequence star located on a nearly circular orbit, highly inclined with respect to the disk, with period $P_{\rm orb}$ of hours to days.

Tidal gravity stretches the stripped stellar debris into an elongated stream subtending approximately $\sim$10\% of the orbital path length before impacting the opposite side of the disk, at approximately $P_{\rm orb}/2$ later. The stream's collision with the rotating disk drives a strong reverse shock which effectively converts the stream's kinetic energy to radiation, producing a bright flare ($L_{\rm bol}\sim 10^{42-43} \, \rm erg \, s^{-1}$) lasting approximately $\Delta t_{\rm s}\sim0.1 \, P_{\rm orb}$. The flare temperature increases with decreasing disk density (or equivalently, lower ejecta mass). At disk densities of order $\rhod \gtrsim 2\times10^{-9} \, {\rm g \, cm^{-3}} \, P_{\rm orb,24}^{7/3}$ the flare peaks in the soft X-ray band, with effective temperatures of $100 -200 \, \rm eV$, compatible with QPE observations.

\subsection{Robust consequences of debris stream-disk interaction}

The emission arising from debris stream-disk collisions naturally depends on the details of several key components, including the ablation of the star's outer layers and the velocity/density distribution of the ejected debris, the flow that develops following stream-disk collision, the accretion disk properties and photon production in the shocked medium. Within the scope of this work, we provided simple analytical approximations, highlighting the appropriate physical scales and the dependencies on key parameters. Numerical experiments and radiation-hydrodynamics simulations are required to validate our assumptions, firm our predictions and to further develop the ideas put forth in this paper. Notwithstanding, we highlight a few key \textit{robust} features of stream-disk interaction, showcasing its potential importance to understanding and interpreting RNTs in general and QPEs in particular.

\begin{enumerate}
    \item \textit{The available energy reservoir}. Stream+disk collisions are ultimately powered by the EMRI orbital energy, $|E_{\rm tot}| \approx m_\star v_{\rm k}^2 /2\approx 10^{52} \, {\rm erg} \, m_1(M_{\bullet,6}/P_{\rm orb,8})^{2/3}$. With every disk passage, ablation liberates a small fraction $m_{\rm ej,\star} \ll m_\star$ of stellar material, whose orbital energy, $\sim m_{\rm ej,\star}v_{\rm k}^2$, is converted to radiation with some efficiency $\varepsilon_{\rm rad}$. The energy reservoir depletion timescale can be related to observable quantities via
    \begin{multline} \label{eq:tau_E_depletion}
        \tau_{\rm E} \approx \frac{P_{\rm orb}}{2} \frac{\varepsilon_{\rm rad}|E_{\rm tot}|}{E_{\rm flare}} \approx \frac{P_{\rm orb}}{2} \frac{\varepsilon_{\rm rad}|E_{\rm tot}|}{L_{\rm bol} \Delta t_{\rm s}} \approx
        150 \, {\rm yr} \\
        \varepsilon_{\rm rad} \pfrac{P_{\rm orb}}{10\,\Delta t_{\rm s}} L_{\rm bol,43}^{-1} \pfrac{M_{\bullet,6}}{P_{\rm orb,8}}^{2/3} m_1 \,.
    \end{multline}
    where $L_{\rm bol,43}=L_{\rm bol}/10^{43} \rm erg \, s^{-1}$. Long-period QPEs with $P_{\rm orb}$ of up to a few days have energetic, long duration flares (e.g., $\Delta t_{\rm QPE} \approx 9 \, \rm hr$ and $E_{\rm flare}\approx 10^{48} \, \rm erg \, s^{-1}$ for AT 2019qiz, \citealt{Nicholl_24}), depleting their energy reservoir on a timescale $\tau_{\rm E} \approx 40 \, {\rm yr} \, \varepsilon_{\rm rad} m_1 (P_{\rm orb}/{3\,\rm d})^{1/3} M_{\bullet,6}^{2/3} (E_{\rm flare}/10^{48} \, \rm erg)^{-1}$. At the other extreme, short period QPEs like eRO-QPE2, have $\tau_{\rm E} \approx 15 \, {\rm yr} \, \varepsilon_{\rm rad} (E_{\rm flare}/10^{46} \, {\rm erg})^{-1} (P_{\rm orb}/5 \, {\rm hr})^{1/3} M_{\bullet,5}^{2/3} m_{0.1}$, normalized to values from \citealt{Arcodia_Linial_24}, and where $m_\star= m_{0.1} \, 0.1\,\rm M_\odot$. The persistence of known QPE sources over the course of several years is consistent with the available orbital energy powering their flares, provided that the radiative efficiency is sufficiently high (i.e., $\varepsilon_{\rm rad} \lesssim 1$).
    
    \item \textit{Stream geometry}. The amount of ablated stellar mass, $m_{\rm ej,\star}$ likely depends on various factors extending beyond those captured in the numerical simulations of \cite{Yao_2025} (e.g., different stellar structures, the gas conditions prevailing in the impacted disk, orbital inclinations etc.). Nonetheless, regardless of the details of the ablation process, the geometry of the debris stream, shaped by the SMBH's tidal gravity can be assessed quite robustly. By solving Hill's equations in the star's frame, we find the debris is stretched to an elongated stream of length $2\ell \sim \mathcal{O}(70) \, r_{\rm H}$ and transverse cross section $A_{\rm s} \approx\pi r_{\rm H}^2$ as it approaches the opposite side of the disk, where $r_{\rm H} = a_0 (m_\star/\MBH)^{1/3}$.

    Importantly, the tidal elongation of the stream sets the flare duration $\Delta t_{\rm s} \approx 2\ell/v_{\rm k} \approx 0.1 \, P_{\rm orb}$ for long period QPEs. Fig.~\ref{fig:TimescaleComparison} demonstrates that for $P_{\rm orb} \gtrsim 1\, {\rm d}$, other emission timescales (e.g., photon diffusion time of shock heated disk material) are considerably shorter than $\Delta t_{\rm s}$, while the stream arrival times appear to match the observations quite well. 
    
    \item \textit{Debris stream blocked by the disk.} 
    The disk mass participating in the stream-disk collision is set by the stream geometry: $m_{\rm s,d} \approx 2\Sigmad \ell r_{\rm H}$. Regardless of specific assumptions about the ablation process, $m_{\rm ej,\star}$ is expected to increase with increasing collision velocity (or shorter $P_{\rm orb}$). For sufficiently long orbital periods, $m_{\rm ej,\star} \ll m_{\rm s,d}$ and the stream is decelerated and deflected by the disk's azimuthal flow, without penetrating through its midplane (see Eq.~\ref{eq:m_ej_vs_m_disk} and accompanying text). We find that this is likely the relevant regime for $P_{\rm orb} \gtrsim$12hr. This behavior is closely related to the interaction of TDE debris streams with AGN disks, as studied by \cite{Chan_2019_tde_agn,Chan_2020_tde_agn,Chan_2021_tde_agn}, who demonstrated that the stream either penetrates or deflected by the disk, depending on the ratio of stream-to-disk mass currents, analogous to the criterion discussed in \S\ref{sec:EjectaStreamCollisionWithDisk}. Unlike the TDE+AGN scenario, the debris streams considered here are short relative to the impact site radius (equivalently, $\Delta t_{\rm s} < P_{\rm orb}$). In TDEs, on the contrary, the fallback debris stream interacts with the disk over the course of many local orbits, strongly perturbing the regions of the disk interior to the stream impact radius, and potentially supressing the SMBH accretion all together. In our case, the limited mass and duration of the stream make such a global impact on the disk highly unlikely.
    
    \item \textit{High radiative efficiency.} A natural consequence of the substantial tidal elongation experienced by the stream is that its total mass is shocked gradually, over a duration $\Delta t_{\rm s}$. This implies that every shocked segment of the stream is locally engulfed by a relatively low optical depth, and thus a substantial fraction of the postshock radiation energy can readily escape, suffering only marginal degradation to adiabatic losses. This process is inherently different from the scenario considered in e.g., \cite{Linial_Metzger_23}, where the entire mass participating in the collision is shocked almost simultaneously and energy deposition is ``explosive''. The high optical depth of the shocked disk material, $\tau \approx \kappa \Sigmad\gg c/v$, implies that the effective radiative efficiency is rather low, scaling as $\varepsilon_{\rm rad} \propto m_{\rm ej}^{-1/2} \ll 1$. As noted before, high radiative efficiencies ($\varepsilon_{\rm rad} \sim 1$) are also supported by the energy output of QPE flares relative to the EMRI orbital energy.

    \item \textit{Soft X-ray emission}. A corollary of high radiative efficiencies $\varepsilon_{\rm rad}\sim 1$, is that the flare temperature closely traces the post-shock temperature, as the emitting region is not obstructed by much excess optical depth. Consider a configuration where sufficient optical depth around the shock front is present to sustain a radiation mediated shock, $\tau \gtrsim c/v_{\rm k}$, and assume that blackbody thermal equilibrium is achieved in the the radiation field downstream of the shock, thus neglecting any potential photon starvation effects.
    Under these assumptions, the postshock temperature is approximately $T_{\rm BB} \approx (\rho_{\rm s} v_{\rm k}^2 / a_{\rm rad})^{1/4}$, where $\rho_{\rm s}$ is the density of the impacting stream. Expressing the stream (debris) mass as $m_{\rm ej,\star} \approx E_{\rm flare}/(\varepsilon_{\rm rad} v_{\rm k}^2)$, and given the stream volume, $A_{\rm s}\ell\approx r_{\rm H}^3 \tilde{\ell}$, we obtain the blackbody temperature
    \begin{multline}
        \kb T_{\rm BB} \approx \frac{\kb}{a_{\rm rad}^{1/4}} \pfrac{E_{\rm flare}}{G m_\star P_{\rm orb}^2}^{1/4} \pfrac{4\pi^2}{\varepsilon_{\rm rad}\tilde{\ell}}^{1/4} \approx \\ \frac{\kb}{a_{\rm rad}^{1/4}}\pfrac{L_{\rm bol}}{G m_\star P_{\rm orb}}^{1/4} \pfrac{4\pi}{\varepsilon_{\rm rad}}^{1/4} \pfrac{m_\star}{\MBH}^{1/12} \approx \\
        170 \, {\rm eV} \, (L_{\rm bol,43}/P_{\rm orb,24})^{-1/4} m_1^{-1/6} M_{\bullet,6}^{-1/12}\,,
    \end{multline}
    where $E_{\rm flare}= L_{\rm bol}\Delta t_{\rm s}$ and $\Delta t_{\rm s}/P_{\rm orb} \approx (m_\star/\MBH)^{1/3}\tilde{\ell}/\pi$. The shock temperature is quite insensitive to the different parameters, and remarkably consistent with the typical QPE observed temperatures, $\kb T_{\rm obs} \approx 100-200 \, {\rm eV}$.

\end{enumerate}

\subsection{Observational Implications}
\begin{enumerate}

    
    \item \textit{Qualitative differences between short and long period QPEs}.
    Recently detected QPE sources \citep{Nicholl_24,Chakraborty_25_22upj,Hernandez_Garcia_25_Ansky} have extended the range of known QPE recurrence times beyond $P_{\rm QPE} \gtrsim 1 \, \rm d$. Based on their long durations and large emitted energies, we argue that in this regime, the observed emission is dominated by the collisions of stellar debris-streams with the underlying accretion disk, rather than the direct EMRI+disk collisions (as proposed in \citealt{Linial_Metzger_23}). At this range of periods, the energy contribution from the direct EMRI+disk collisions is generally outweighed by the stream component (e.g., tables \ref{tab:inner_disk}, \ref{tab:outer_disk}).

    For sufficiently short-period QPEs ($\left<P_{\rm QPE}\right> \approx {\rm few}\, \rm hr$), the emission arising from the direct interaction of the EMRI's Hill sphere (filled with ablated stellar material), may be energetically comparable to the stream+disk interaction. Which of these components dominates the observed signal depends on their emission temperature, potentially governed by effects of photon starvation in the shocked medium and deviations from thermal equilibrium \citep{Vurm_25}. Furthermore, we find that for sufficiently short orbital periods, ejecta streams may easily penetrate through the disk, effectively increasing the collision cross section. In this regime, the emission likely resembles the predictions of the shocked-disk picture, with effective cross sections that may greatly exceed the star's size (see also \citealt{Vurm_25} in this context). In this regime, the duration of flares is set by the stream crossing time through the disk, and the radiating mass is the shocked disk mass, $m_{\rm s,d}$ (Eq.~\ref{eq:m_shocked_disk_mass}). 
    
    \item \textit{One versus two flares per orbit}. 
    We describe two important emission sources: (1) Cooling emission of shocked disk material, carved out either directly by the star, or by the relatively dense stellar debris, and (2) Cooling emission of shocked stellar debris, following its impact with comparatively denser disk.

    When a star passes through a disk with $\Hd \gtrsim R_\star$, two plumes of shock heated disk material are ejected above and below the plane of the disk, as seen in numerical simulations (e.g., \citealt{Ivanov_98} and \citealt{Huang_2025}). An important consequence of the presence of dual ejecta clouds, is that both ingress and egress disk passages produce bright (and likely comparable) emission, such that two similar flares are visible with every orbit for every observer position. Indeed, the observed pattern of alternating long/short recurrence times in some QPE sources (eRO-QPE2, RXJ+1301, GSN 069) is naturally explained by an EMRI on a mildly eccentric orbit, producing a visible flare with every disk passage.

    On the contrary, the emission from a low-mass debris stream colliding with the disk is confined to one side of the disk at a time, if the stream does not have sufficient inertia to penetrate through the disk. While the opposite side of the disk is perturbed by the impact, stream+disk interaction is more likely to produce just one visible flare per orbit (as the star moves away from the observer and into the disk). Two flares per orbit are possible if the disk is viewed edge-on, or if the ejecta stream is sufficiently massive to puncture through the disk (see Eq.~\ref{eq:m_ej_vs_m_disk}).

    \item \textit{Flare variability}. While some QPEs follow a relatively regular timing pattern, other sources appear to be more erratic: eRO-QPE1 demonstrates complex flare structure and timing, potentially composed of overlapping emission components \citep{Arcodia_2022_complex}; AT 2022upj appears to sporadically miss flares \citep{Chakraborty_25_22upj}. Although apsidal/nodal orbital precession and/or rapid disk precession may account for some of the timing phenomenology \citep[e.g.,][]{Xian_2021,Franchini_23,Chakraborty_2024,Zhou_2024}, other mechanisms may be at play. Variations in the amount of ablated stellar mass between different disk passages, fluctuations in the disk condition encountered by the EMRI or perturbations to the stream trajectory may all manifest as sources of variability in the X-ray lightcurve. Since the collision duration is set by the stream length, which is susceptible to various perturbative effects, flare timing should not be directly interpreted as a probe of the EMRI orbit, since, e.g., it is unclear that time of peak X-ray flux actually traces the EMRI's disk crossing.  Improved understanding of stream-disk collision dynamics and the resulting lightcurves would allow for more accurate orbital modeling based on QPE observations \citep[e.g.,][]{Xian_2021,Franchini_23,Zhou_2024}. 

    Particularly erratic behavior may arise when the debris mass is comparable to the impacted disk mass (i.e., when $m_{\rm ej,\star} \gtrsim m_{\rm s,d}$). In this regime, the hydrodynamics of the stream-disk collision is particularly messy -- streams may partly penetrate through the disk, resulting in an outflow quite distinct than that envisioned in \S \ref{sec:Observables} (and Fig.~\ref{fig:Cartoon_Star_Stream_Disk_Interaction}). In particular, previously ablated material may survive a few disk collisions before considerable deceleration. Multiple fragments of debris streams from previous passages may be present, impacting the disk and producing a complex, multi-flare signal, possibly reminiscent of that seen in eRO-QPE1.

\end{enumerate}

\subsection{Detectability Criteria of X-ray QPEs} \label{sec:QPE_detectability}
Our model for stream-disk encounters suggests that QPE-like soft X-ray flares arise when a radiation-mediated shock forms, requiring an optical depth $\tau_{\rm sh} \gtrsim c/v$. Under the assumptions of the ablation model of \cite{Yao_2025}, the underlying disk density must therefore be comparable to $\rhod^{\varepsilon_{\rm rad} \approx 1} \approx 2.5 \times10^{-8} \, {\rm g \, cm^{-3}} \, P_{\rm orb,24}^{7/3}$ (Eq.~\ref{eq:rho_eff_1}), such that for $\rhod > \rhod^{\varepsilon_{\rm rad} \approx 1}$, a radiation mediated shock develops at the stream-disk interface. At substantially higher densities, the resulting flare temperature peaks in softer bands (UV/optical), and it is likely obscured by the bright emission from the disk. We note, however, that at sufficiently soft bands, the flare emission may outshine the low-frequency tail of the disk spectrum, similar to the scenario studied in \cite{Linial_Metzger_24a}. When this is the case, repeating UV flares (``UV-QPEs''), analogous to the population of X-ray QPEs may be detectable. At low disk densities, $\rhod \ll \rhod^{\varepsilon_{\rm rad} \approx 1}$ the shock is collisionless, and we expect it to be radiatively inefficient, producing a signal that is harder than the observed QPE range (possibly hard X-rays or even gamma rays), and bolometrically fainter.

As the underlying TDE spreads and its accretion rate decreases, a wide range of densities $\rhod$ is naturally realized at a fixed radius. This suggests that soft X-ray QPEs ``turn on'' when the disk density at $a_0$ approaches $\rhod^{\varepsilon_{\rm rad} \approx 1}$. If the EMRI is initially well outside the disk, collisions commence at time $t_{\rm disk}(a_0)$, when $R_{\rm d}\approx a_0$. The local density encountered by the EMRI then increases over a timescale comparable to $t_{\rm disk}(a_0)$, spending a time $\eta_\Sigma^{\rm X} t_{\rm disk}$ with visible soft X-rays. 

A spreading TDE disk model \citep{Mummery_2024} fitted to the quiescent optical/UV/X-ray emission in AT 2019qiz and AT 2022upj demonstrates that the disk's outer radius is comparable to the semi-major axis of an EMRI with $P_{\rm orb} = \left< P_{\rm QPE} \right>$ \citep{Nicholl_24,Chakraborty_25_22upj}. This implies that currently, $\eta_\Sigma \approx 1$ for both of these sources, and thus X-ray QPEs are expected to persist over a timescale comparable to the disk's current spreading time, $t_{\rm disk}$, before the local density evolves significantly. It is worth noting that in AT 2022upj, no QPEs were observed during the initial late-time rise of X-ray flux (time $t_1$), when the fitted disk model shows $\Sigmad(a_0,t_1)/\Sigmad(a_0,t_2) \approx 10^{-3}$, where $t_2$ is the epoch at which QPEs were first detected. This could be interpreted as the absence of enough stream mass to produce an RMS at this early stage.

Lastly, the surface density values implied by these disk models, $\Sigmad \approx 10^3 \, \rm g \, cm^{-2}$ yield midplane densities, $\Sigmad/\Hd \approx 2\times 10^{-9} \, (\Sigmad/10^3 \, {\rm g \, cm^{-2}}) M_{\bullet,6}^{-1/3} (P_{\rm orb}/2\,{\rm d})^{-2/3} \tilde{h}_{-2}^{-1}$ that are too low (by at least an order of magnitude) relative to the predicted $\rho_{\rm d}^{\varepsilon_{\rm rad}\approx 1}$ (Eq.~\ref{eq:rho_eff_1}) for the assumed fiducial values. We note, however, that the overall normalizations of the surface-density profiles implied by fitting of disk models is not well constrained, and higher disk densities are also permitted. Alternatively, the critical disk density for which $\varepsilon_{\rm rad} \sim 1$ may be lower if the ablation efficiencies of EMRIs in these sources are higher than we assumed, or if mass is removed from the EMRIs by another mechanism, e.g., Roche lobe overflow at pericenter passage \citep[e.g.,][]{Krolik_Linial_2022,Linial_Sari_2023,Lu_Quataert_23,Yao_2025b}. This latter possibility would imply shorter ablation timescales, and cannot fully compensate for low values of $\rhod$.

In two of the short-period QPEs, GSN 069 and eRO-QPE2, fitting of the joint UV/X-ray quiescent emission suggests disks that extend well-beyond the EMRI's orbit \citep{Wevers_2025_ero2,Guolo_2025_GSN069_joint}. The density encountered by the EMRIs varies with time as the disk continues to spread, and its accretion rate $\dot{M}_{\rm d}$ drops. While the overall mass of the spreading disk decreases with time, the mass density encountered at a given location may either increase or decrease, depending on the disk scale height $H_{\rm d}(t;a_0)$ evolution as the disk continues to cool, relative to the evolution in $\Sigmad(t;a_0)$. A secular increase in the local density seen by the EMRI present in GSN 069 may explain why X-ray QPE flares were absent in GSN 069 in 2014 but were seen in 2018: lower mass ejecta (arising from low $\rhod$) in 2014 resulted in a radiatively inefficient collisionless shock at the stream-disk interface. After the density increased to roughly $\rhod \approx \rhod^{\varepsilon_{\rm rad}\approx1}$ the collisions became radiatively efficient, peaking in the soft X-rays. Alternatively, if the local density has decreased with time, the flares produced by EMRI collisions in 2014 may have been too soft to overshine the disk emission. As the disk density continued to decrease, the resulting stream became more dilute (Eq.~\ref{eq:rho_s_relative}), such that photon starvation effects became important by 2018, rendering the effective emission temperature harder, producing visible X-ray flares at this epoch and later (see \citealt{Guolo_2025_GSN069_time_dependent} for further discussion of this possibility).

\subsection{Caveats and future directions}

We estimated the resulting emission of various components of the EMRI-disk interaction (direct encounters between the `bare' star or the dense ejecta within its Hill sphere and the disk, as well as the elongated stream-disk collisions) in \S \ref{sec:Observables}. These derivations relied on the assumption that photon production and their thermalization in the shocked downstream is rapid, and that the resulting shocks are radiation mediated.

These assumptions certainly fail in parts of parameter space, e.g., when the stream is too dilute, and the optical depth near the shock front falls below $\tau_{\rm sh} \lesssim c/v$ (Fig.~\ref{fig:M_ej_Period_Diagram}). We did not explore here the details of the emission at disk densities marginally below $\rhod^{\varepsilon_{\rm rad} \approx 1}$, where shocks are expected to be collisionless. Various effects, including potential inverse-Compton scattering of the soft emission from the underlying disk (e.g., as explored in \citealt{Chan_2021_tde_agn}), cooling by free-free emission and Comptonization of the downstream radiation will be important for predicting the observational signatures of this regime. One possible outcome of low-density streams impacting the disk, is the existence of repeating, hard X-ray/gamma-ray flares, in a phase predating the onset (or preceding) of soft X-ray QPEs (for example, in the limit $a_0\approx R_{\rm d}$ this may occur earlier in the rise of the disk density encountered by the EMRI, before soft X-ray QPEs appear). We are unaware of any current observations that confirm or rule out this possibility.

A potential shortcoming of our model is that it fails to account for the long-term fate of the ablated stellar material. One possibility is that the excess mass removed from the star is effectively entrained in the accretion flow feeding the black hole. Alternatively, a fraction of the shocked stream material may accumulate to engulf the entire system with ever-increasing optical depth (akin to the scenario laid out in appendix C of \citealt{Linial_Metzger_23}). Given the ablation mass loss rate (Eq.~\ref{eq:mdot_abl}) and the disk densities conducive for X-ray flares (e.g., Eq.~\ref{eq:rho_eff_1} and Fig.~\ref{fig:M_ej_Period_Diagram}), mass is added to the system at a rate $\dot{m}_{\rm abl} \approx 10^{-2} \, \dot{M}_{\rm Edd} \, P_{\rm orb,8}^{-5/3}$, which may be comparable to the disk accretion rate in sources of short period. If a substantial fraction of this mass is added to the disk, its evolution may be modified due to the additional mass and energy sourced from the interacting EMRI. In the extreme limit where an $\mathcal{O}(1)$ fraction of the ablated mass engulfs the EMRI orbit, remaining hot and highly ionized, the scattering optical depth engulfing the system exceeds unity after a mass of merely $m_{\rm \tau=1} \approx 4\pi a_0^2 / \kappa_{\rm es} \approx 3.7\times 10^{-6} \, {\rm M_\odot} M_{\bullet,6}^{2/3} P_{\rm orb,8}^{4/3}$ accumulates. With disk densities implied by Eq.~\ref{eq:rho_eff_1}, such mass is shed from the star within merely $\lesssim 10 \, P_{\rm orb,8}^{2/3}$ orbits (or $62 \, {\rm hr} \, P_{\rm orb,8}^{5/3}$). Since the observations do not show signs of significant reprocessing of the flare/quiescent emission, it appears that the shocked ablated mass does not ultimately enshroud the entire system. Hydrodynamic star-disk collision simulations may provide clues regarding the long term fate of the shocked debris/disk material powering the observed flares and its observational implications.

We assumed that the EMRI follows a circular orbit, and derived the stream geometry and evolution under this assumption (e.g., in the form of Hill's equations in \S \ref{sec:Debris_and_stream}). Dynamical modeling of QPE timing patterns have been used to constrain the EMRI orbital eccentricities, finding $e \lesssim 0.1$ or smaller in most sources \citep{Zhou_2024}. Our analytical estimates of the resulting stream-disk interaction could be generalized to arbitrary eccentricities, where different stream densities and collisions durations ($\Delta t_{\rm s}$) are expected. In particular, our results could be applied to other classes of RNTs, such as ASASSN-14ko \citep{Payne_2021}, where interaction between tidal streams from a partially disrupted star on an eccentric orbit ($e \gtrsim0.95$), and an AGN accretion disk is believed to take place \citep{Linial_Quataert_24a}. More broadly, EMRIs with high orbital eccentricities ($e\gtrsim0.5$)  \citep[e.g.,][]{Linial_Sari_2023}, could have similarly high-velocity disk collisions (near pericenter) as short-period, circular EMRIs (where $(1-e)a_0$ replaces $a_0$ throughout many of the expressions along the paper). Thus, two sources of markedly different recurrence times may show similar flare energetics, provided that their pericenter distances are comparable (see section 4.2 of \citealt{Linial_Metzger_23} for further discussion of this point).

\begin{acknowledgments}
IL would like to thank Yuri Levin, Riccardo Arcodia, Erin Kara, Joheen Chakraborty, Philippe Yao and Re'em Sari for stimulating discussions which have culminated in this paper. IL also acknowledges support from a Rothschild Fellowship and The Gruber Foundation. This research benefited from interactions at workshops funded by the Gordon and Betty Moore Foundation through grant GBMF5076. This research was partially funded by NASA grant 80NSSC24K0934. BDM was supported in part by the Simons Foundation (grant 727700 and 827103). The Flatiron Institute is supported by the Simons Foundation.
\end{acknowledgments}

\bibliography{main}{}
\bibliographystyle{aasjournal}

\end{document}